\documentclass{sig-alternate}

\begin{document}
\conferenceinfo{ICDCN}{2017}

\title{Total Order Reliable Convergecast in WBAN}
 
\numberofauthors{2} 

\author{ 
\alignauthor
Gewu Bu\titlenote{This work has been supported by the SMARTBAN project, Labex SMART}\\
       \affaddr{Universit\'e Pierre et Marie Curie, LIP6, CNRS UMR 7606}\\
       \affaddr{4, place Jussieu, Paris, France}\\
       \email{ gewu.bu@lip6.fr}
% 2nd. author
\alignauthor
Maria Potop-Butucaru\\
       \affaddr{ Universit\'e Pierre et Marie Curie, LIP6, CNRS UMR 7606}\\
       \affaddr{4, place Jussieu, Paris, France}\\
       \email{maria.potop-butucaru@lip6.fr}
}
 
\date{ }

\maketitle
\begin{abstract}
This paper is the first extensive work on total order reliable convergecast in multi-hop Wireless Body Area Networks (WBAN). Convergecast is a many-to-one cooperative scheme where each node of the network transmits data towards the same sink. Our contribution is threefold. First, we stress existing WBAN convergecast strategies  with respect to their capacity to be \emph{reliable} and to ensure the \emph{total order} delivery at sink. That is, packets sent in a specific order should be received in the same order by the sink. When stressed with transmission rates up to 500 packets per second the performances of these strategies decrease dramatically (more than 90\% of packets lost). Secondly, we propose a new posture-centric model for WBAN. This model offers a good characterization of the path availability which is further used to fine tune the retransmission rate thresholds. Third, based on our model we propose a \emph{new mechanism for reliability} and a \emph{new converge-cast strategy} that outperforms  WBAN dedicated strategies but also strategies adapted from DTN and WSN areas. Our extensive performance evaluations use essential parameters for WBAN: packet lost, total order reliability (messages sent in a specific order should be delivered in that specific order) and various human body postures. In particular, our strategy ensures zero packet order inversions for various transmission rates and mobility postures. Interestingly, our strategy respects this property without the need of additional energy-guzzler mechanisms. 
\end{abstract}

%\begin{IEEEkeywords}
%Wireless Body Area Networks, Total Order Reliable Convergecast, Networks Modelization
%\end{IEEEkeywords}

%\IEEEpeerreviewmaketitle

\section{Introduction}
%\blindtext
Wireless Body Area Networks (WBAN) is a cross-area between Wireless Sensor Networks (WSN) and Delay Tolerant Networks (DTN) with as main objective the collection of physiological parameters from sensors deployed on a human body that undergo the human mobility.  Designing efficient protocols for this area is a challenging task. One of the particularities of the WBAN area are the specific rates of the physiological parameters monitoring. In \cite{khan2010wireless} the authors advocate that in the current applications of WBAN, the transmission rate is less than 10 packets per second. However, these rates may drastically increase in the future. Therefore, in order to meet the requirements of medical monitoring, WBAN is needed to withstand a continuous packets flows. In some critical applications such as monitoring patients during a surgery or post-surgery the flows carry vital alerts. Missing some of them or wrongly interpret the flow of data due to packets inversions may have a tremendous impact. In this context designing efficient total order reliable protocols is crucial for saving human lives.   

In this paper we focus the \emph{convergecast} communication primitive since this is one of the main building block in WBAN. Convergecast allows nodes to transmit data towards a sink.  Convergecast has the total order property if packets are delivered at the sink in the same order as they have been transmitted.

%\subsubsection{Fiability and Total Order in WBAN}
Recently, \cite{badreddine2015convergecast} surveys the existing work on convergecast on various areas including Delay Tolerant Networks, DTN and Wireless Sensor Networks, WSN. They argue that most of the existing strategies are not directly implementable in WBANs due to their needs in computing capacities, memory or their energy consumption. Finally, building on top of \cite{tseng2002broadcast, groverreliable, karthikeyan2013comparative, sharma2014performance, rehena2011modified}, they propose and evaluate three classes of convergecast strategies taylored for WABN: 1) Multi-Paths based strategies, 2) Attenuation-based strategies and 3) Gossip-based strategies. Their evaluation does not target the total order delivery property of these strategies neither their resistance to increasing flows of messages. It merely focus on the resilience to the human body mobility and energy consumption. They measure the following parameters: percentage of received messages (under the hypothesis that each node sends a single message in each run), end-to-end delay and the number of transmission/receptions since this is a good indicator of the energy consumption.

In terms of \emph{reliability}, to the best of our knowledge there are three main reliability mechanisms used in WBAN. The first mechanism computes and presets a static or dynamic overlay path for collecting data. Strategies proposed in \cite{kim2016ahp, zhang2015reliable} and attenuation-based strategies in \cite{badreddine2015convergecast} choose a reliable leader or a set of relay nodes as the next hop(s) to help in forwarding the packets. Strategies proposed in \cite{colesanti2011collection, ghadimi2014opportunistic, adhikary2016new, kachroo2015novel, kaur2015cost,  juneja2015reliable, abd2015self, juneja2015tree} care and learn in the entire network paths by sending BEACON or HELLO message in order to construct a path from every source to the sink. These paths are then dynamically updated during the packets sending. The Multi-Paths based strategies in \cite{badreddine2015convergecast} use a specific static path to collect data.

The second mechanism use special retransmission techniques at MAC level \cite{deepak2016enhancing} or a TCP-like reliability flow control mechanism at Transport level \cite{kathuria2015reliable}. However, they are more like Cross-Layer strategies \cite{karvonen2015cross, tseng2015efficient}. In this work we specifically address only the network level strategies.

The third mechanism uses either compressive sensing (CS) technology to reconstruct missing packets \cite{liu2015cs} conservation central network CCN-based model \cite{quan2015wban}. This third mechanism is particular energy-guzzler.

In terms of \emph{total order} (packets delivery that respect the sending order), to the best of our knowledge no extensive study has been conducted so far in WBAN.

%\subsection{Our contribution} 
\paragraph{Our Contribution} Our work on multi-hop WBAN was motivated by the recent findings in \cite{naganawa2015simulation} where it is proven that multi-hop communication has better performances face to human body mobility than classical on hop communication. Our work is the first to extensively study the total order reliability of convergecast in multi-hop WBAN. Our contribution is \emph{threefold}. \emph{First} we simulate and analyse  thirteen representative strategies for convergecast some of them tailored for WBAN others borrowed from the sensor network or DTN literature. We stress them with various packets rate up to 500 packets per second. We evaluate the throughput as well as their capability to totally order deliver packets (packets are delivered at the sink in the same order as they have been transmitted). 
We performed our evaluations with OMNeT++ simulator that we enriched with realistic human body mobility and channel models issued from the recent research on biomedical and health informatics \cite{naganawa2015simulation}. 
Our simulations show that the throughput is inversely proportional with the increase of transmission rate. The packets loss goes to 90\%. This first contribution is of independent interest since it lays the bases for dining efficient convergecast strategies in multi-hop WBAN.
Our \emph{second contribution} is a \emph{new model for WBAN} and based on this new model we propose a \emph{new mechanism} for increasing the reliability of convergecast.  Interestingly, equipped with this new mechanism convergecast strategies improve their performances both in terms of throughput and total order reliability for a packet rate that goes as far as 20 packets per second. Note that this threshold is twice bigger than the maximum packet rate of actual WBAN applications \cite{khan2010wireless}. However, this mechanism does not help in avoiding sequence inversions. Our \emph{third contribution} is a new convergecast strategy that ensures 100\% total order reliability.  

\paragraph{Roadmap}
In Section \ref{existing} we simulate and analyse thirteen different strategies: four Multi-Paths based strategies \cite{badreddine2015convergecast} as examples of the presetting reliable overlay paths; four Attenuation-based strategies \cite{badreddine2015convergecast} as examples of reliable choice of the next-hop; the Collecting Tree Protocol (CTP) \cite{colesanti2011collection} and the Opportunistic Routing (ORW) \cite{ghadimi2014opportunistic} as examples of entire network reliable strategies. Furthermore, we analyse three Gossip-based strategies \cite{badreddine2015convergecast} as example of flooding-based strategies. In Section \ref{model} we present our new model for WBAN. In Section \ref{mechanism} we introduce our new reliability mechanism. Finally, in Section \ref{strategy} we introduce our new convergecast strategy that out-performs existing ones in terms of total order reliability.

\section{Analysis of existing Convergecast Strategies}
\label{existing}
In this section, we stress strategies discuss in \cite{badreddine2015convergecast} plus the two reference strategies CTP \cite{colesanti2011collection} and ORW \cite{ghadimi2014opportunistic}. More specifically, we analyse throughput when the transmission rate goes as far as 500 packets per second. Our simulations prove that their performances (without exception) proportionally decrease with the increase of the rate transmission. 

In the sequel we present briefly the channel model we used, then the simulation environment. Furthermore, we briefly present each of the strategies we simulated and finally the simulation results. 

\paragraph{Channel Model and Simulation Environment}
%In the area of WBAN, we inevitably have to mention the mobility of each WBAN sensor, we call nodes, witch move with the movement of their wearer. WBAN sensors have special characteristics as the low transmission power, the short transmission range and the low computing and memory capacity. Due to these characteristics, the transmission range in WBAN for each node is comparable with human body size. It means the problematic of mobility in WABN is no more a broad 2D plan like traditional Wireless Sensor Networks but a small 3D space network. That makes the study of WBAN mobility is more complicated and makes some current research for 2D topology for WSN\cite{silva2012proposal,das2015energy,islam2016mobility} ineffective.

\begin{figure}
\centering
\epsfig{file=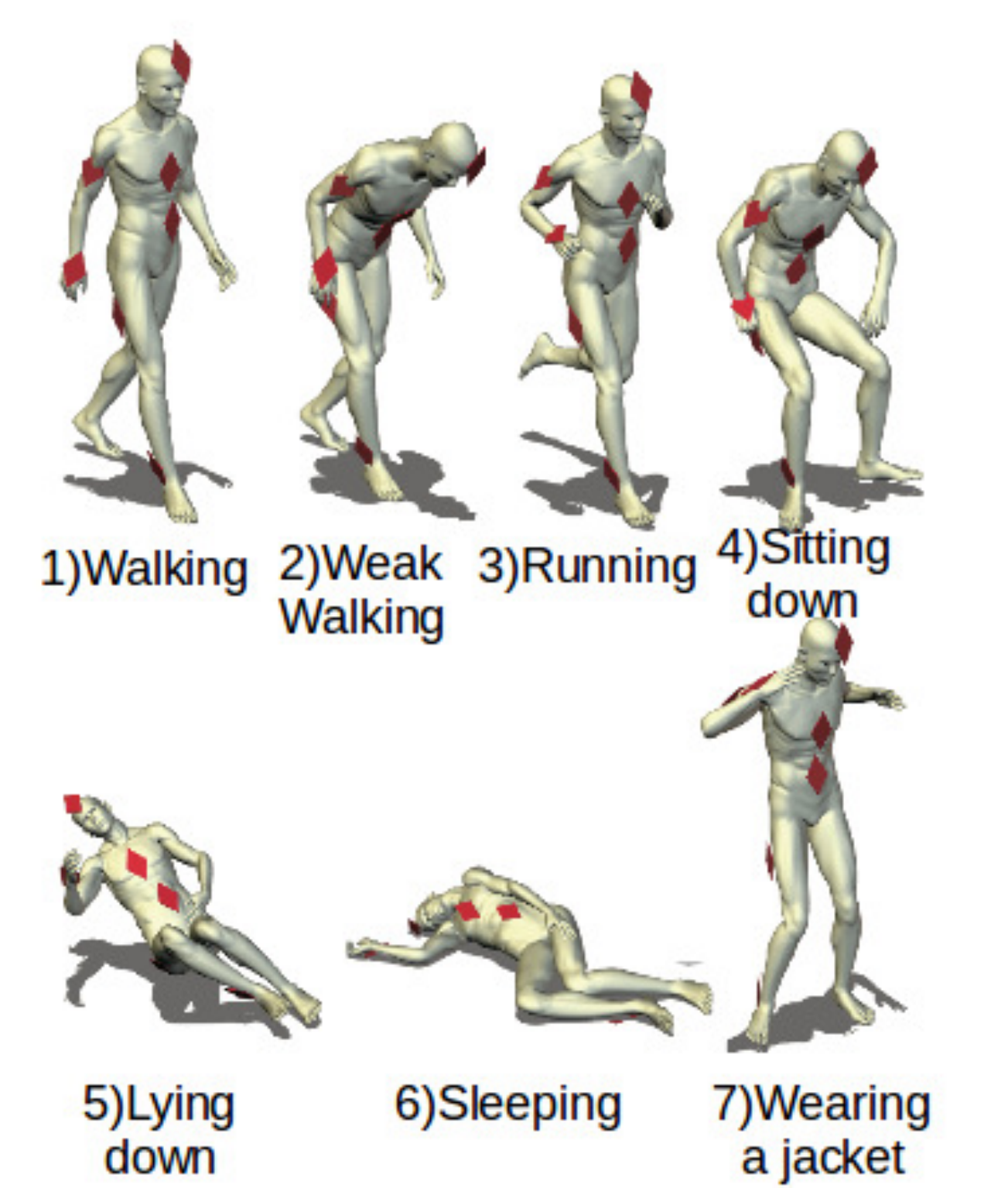, width=2in}
\caption{Seven postures for different human mobility}
\label{Figposture}
\end{figure}

%From recent research for channel model and mobility model in WBAN, most proposed models give a single valuer as the modeling characteristic: the signal strength attenuation. Normally attenuation valuers are calculated by the continuous measuring of moving person who wears several WABN sensors on his body\cite{wangchuk2013cooperator}.
%Two channel models have been proposed so far in WBAN \cite{wangchuk2013cooperator} and \cite{naganawa2015simulation}. The model in \cite{wangchuk2013cooperator} use the signal strength attenuation due to the low transmission power, wireless signal can be attenuated obviously by human body and environment, like the reflection, the refraction, the absorb and even clothes occlusion. The attenuation values are calculated by the continuous measuring of moving person who wears several WBAN sensors on their body.
In this work we use the channel mobility model proposed in  %\cite{badreddine2015broadcast} based on the data sets of 
\cite{naganawa2015simulation}. In \cite{naganawa2015simulation} the authors proposed a simulate-based channel model based on mobility data set for WBAN issued from experiments with a network composed of seven sensors distributed on the body as follows 
%(see Figure \ref{Fig1)})
: 0)navel, 1)chest, 2)head, 3)upper arm, 4)ankle, 5)thigh and 6)wrist. Using a software-simulated-human-body model to replace the real person,  the authors measure the mean and the standard deviation of the channel attenuation between every two nodes in seven different postures: 1)Walking, 2)Walking weakly, 3)Running, 4)Sitting down, 5)Lying down, 6)Sleeping and 7)Wearing a jacket, respectively  (see Figure \ref{Figposture}).  

Furthermore, \cite{naganawa2015simulation} advocate that multi-hop communication achieves better performances than one-hop star topology considered so far in WBAN.
 Based on this data set, \cite{badreddine2015broadcast}  proposed a new channel-mobility: for every packet sent from a source, a random attenuation will be added. If the signal strength after adding the attenuation is smaller than the sensibility at the receiver, the packet will be dropped. Each two nodes have different attenuation between them: the random attenuation is calculated by the specific mean and standard deviation according to one of pairs source and destination from the normal distribution. 

%The advantage of this simulate-date-set-based channel-mobility model has the smaller individual differences comparing with the measuring method\cite{wangchuk2013cooperator}, because that the height, posture, exercise habits of the tested person and even external environment interference will be very variant in different measuring-based cases.  

%\begin{figure}[!t]
%\label{Fig1}
%\centering
%\includegraphics[width=.4\columnwidth]{posture.png}\\
%\caption{WBAN Nodes on the Body}
%\label{ExecutionTime}
%\end{figure}

%\paragraph{Simulation Environment}
%In this paper, we use the OMNET++ simulator enriched with the  MIXIM project and the channel-mobility model of \cite{badreddine2015broadcast}. The 
%network topology we consider is shown in Figure \ref{Fig1} where the 
% sink node is node 1.
In this paper, we use the same simulation environment as in \cite{badreddine2015broadcast} (IEEE 802.15.4) with a communication frequency of 2.45 GHz. 
%\begin{itemize}
%\item {Simulator}: OMNET++ 4.6 and MIXIM 2.3 framework.  
%\item {Communication frequency}: 2.45GHz.
%\item {MAC Protocol}: CSMA/CA. 
%\end{itemize}

The transmission power and the sensibility of the radio module of nodes are set to -60dBm and -100dBm respectively. A random channel attenuation for every transmission between different nodes given by a realist channel model to represent intermittent links in a realistic environment and also the mobility of nodes on human-body.

\subsection{Description of analyzed  strategies}
Convergecast is a many-to-one cooperative scheme where each node of the network  transmits data towards the same sink.
In \cite{badreddine2015convergecast} the authors survey and classify existing convergecast strategies for WBAN. We simulated these strategies altogether with CTP \cite{colesanti2011collection} and ORW \cite{ghadimi2014opportunistic} on top of a network composed of seven nodes using a channel and body mobility model as described in the previous section.
 
In the sequel we present briefly the simulated strategies followed by the detailed presentation of the simulation results.
As explained previously, we consider multi-hop communication and the human mobility model presented in \cite{naganawa2015simulation} Figure \ref{Figposture}.
\emph{Multi-Paths based strategies} %(Figure \ref{Fig2})
: these strategies are based on predetermined paths and use this pre-set overlay as a reliability mechanism.  
\begin{itemize}
\item {APAP}: Each node sends or forwards packets to all its parents until packets reach the sink.  
\item {APPP}: Each source node sends its packets to all its parents and parents randomly forward packets to one of their parents.
\item {PPAP}: Each source node randomly sends its packets to one of its parents and parents forward packets to all their parents.
\item {PPPP}: Each node randomly sends or forwards packets to one of its parents.
\end{itemize}

%\begin{figure}[!t]
%\centering
%\includegraphics[width=.9\columnwidth]{picture1/newV2/APAP.png}

%\centering
%\includegraphics[width=.9\columnwidth]{picture1/newV2/APPP.png}

%\centering
%\includegraphics[width=.9\columnwidth]{picture1/newV2/PPAP.png}

%\centering
%\includegraphics[width=.9\columnwidth]{picture1/newV2/PPPP.png}
%\caption{Multi-Paths based strategies}
%\label{Fig2}
%\end{figure}

%\begin{figure}[!t]
%\centering
%\includegraphics[width=.9\columnwidth]{picture1/newV3/newAPAP.png}
%\caption{Multi-Paths based strategies}
%\label{Fig2}
%\end{figure}

\emph{Attenuation-based} strategies 
%(Figure \ref{Fig3})
: these strategies are based on the negotiation of channel attenuation.  When a source has packets to send, it broadcasts firstly a Request to ask an estimated attenuation from the receiver of this Request to the sink then the receiver of this Request will send back a Reply with the required attenuation value. The source receiving Replies will chose a next hop among replying nodes and sends data packets to them; if no Reply has been received for a while, the source will re-send a Request.
\begin{itemize}
\item {MinAtt}: Each source chooses as next hop the nodes with  the best (smallest) attenuation value among the responders.
\item {BothMinAtt}:
Each source chooses as next hop two nodes who give the two best (smallest) attenuation values  among all the responders.
\item {CloseToMe}:
The source requires not only the attenuation value to the sink but also the attenuation to the source and chooses two nodes: the ones  that give the two best (smallest) attenuation valuers, according to who has a smaller attenuation value to the source.
\item {RandAtt}:
The source randomly chooses one responder as the next hop.
\end{itemize}

%\begin{figure}[!t]
%\centering
%\includegraphics[width=.9\columnwidth]{picture1/newV2/MinAtt.png}

%\centering
%\includegraphics[width=.9\columnwidth]{picture1/newV2/BothMinAtt.png}

%\centering
%\includegraphics[width=.9\columnwidth]{picture1/newV2/CloseToMe.png}

%\centering
%\includegraphics[width=.9\columnwidth]{picture1/newV2/RandAtt.png}
%\caption{Attenuation-based Strategies}
%\label{Fig3}
%\end{figure}

%\begin{figure}[!t]
%\centering
%\includegraphics[width=.9\columnwidth]{picture1/newV3/newAtt.png}
%\caption{Attenuation based strategies}
%\label{Fig3}
%\end{figure}
%
\emph{Dynamic Path Strategies} 
%(see Figure \ref{Fig4})
: these strategies construct and update an overlay rout over time.

\begin{itemize}
\item{CTP} All the sources periodically broadcast BEACON messages for constructing and updating a network-cost metric, according to which each node chooses his next hop to the sink.
\item{ORW} Nodes get their  parents in an initial phase where PROBING and REPROBING messages are exchanged. After this initial phase, nodes update the network-cost metric by computing data packets's receiving rate on each node. Sources send packets to a set of relay nodes.
\end{itemize}

%\begin{figure}[!t]
%\centering
%\includegraphics[width=.9\columnwidth]{picture1/newV2/CTP.png}

%\centering
%\includegraphics[width=.9\columnwidth]{picture1/newV2/ORW.png}
%\caption{Dynamic Path Strategies}
%\label{Fig4}
%\end{figure}

%\begin{figure}[!t]
%\centering
%\includegraphics[width=.9\columnwidth]{picture1/newV3/newCO.png}
%\caption{Dynamic Path based strategies}
%\label{Fig4}
%\end{figure}

\emph{Gossip-based} strategies 
%(see Figure \ref{Fig5})
: these strategies are based on packets flooding.

\begin{itemize}
\item{FloodToSink}:
Nodes broadcast their own packets and the received packets to their neighbours until packets reach the sink or the limitation of TTL reached.
\item{ProbaCvg}:
Nodes broadcast their own packets and the packets received to their neighbours with a probability until packets reach the sink or the limitation of TTL reached. Each time of broadcast will halve the forwarding probability.
\item{PrunedCvg}:
Nodes send their own packets and the packets received uni-cast to a random node as next hop.
\end{itemize}

%\begin{figure}[!t]
%\centering
%\includegraphics[width=.9\columnwidth]{picture1/newV2/FllodToSink.png}

%\centering
%\includegraphics[width=.9\columnwidth]{picture1/newV2/ProbaCvg.png}

%\centering
%\includegraphics[width=.9\columnwidth]{picture1/newV2/PrunedCvg.png}
%\caption{Gossip-based Strategies}
%\label{Fig5}
%\end{figure}

%\begin{figure}[!t]
%\centering
%\includegraphics[width=.9\columnwidth]{picture1/newV3/newCvg.png}
%\caption{Gossip-based strategies}
%\label{Fig5}
%\end{figure}

\subsection{Simulation Results}
%We stressed the above described strategies with various transmission rates in order to  meet the requirements of realistic medical applications \cite{khan2010wireless}.  
%We focus their throughput in a realistic environment of intermittent links when the transmission rate goes as far as 500 packets per second and their ability to deliver packets respecting the total order property in context of WBAN.  
%However, instead of using only one packet sent by the source nodes  as in the simulation results proposed in  \cite{badreddine2015broadcast} and \cite{badreddine2015convergecast}) we considered  
Due to the lack of space, in this section we present solely the results for the \emph{Walking} posture. We measure the percentage of received packets by the sink when the packets rate varies from 1 packet to 500 packets per second. Furthermore, we address the total order property.
%\subsection{Total order Reliability Analysis}

\subsubsection{Throughput}
\begin{figure}
\centering
\epsfig{file=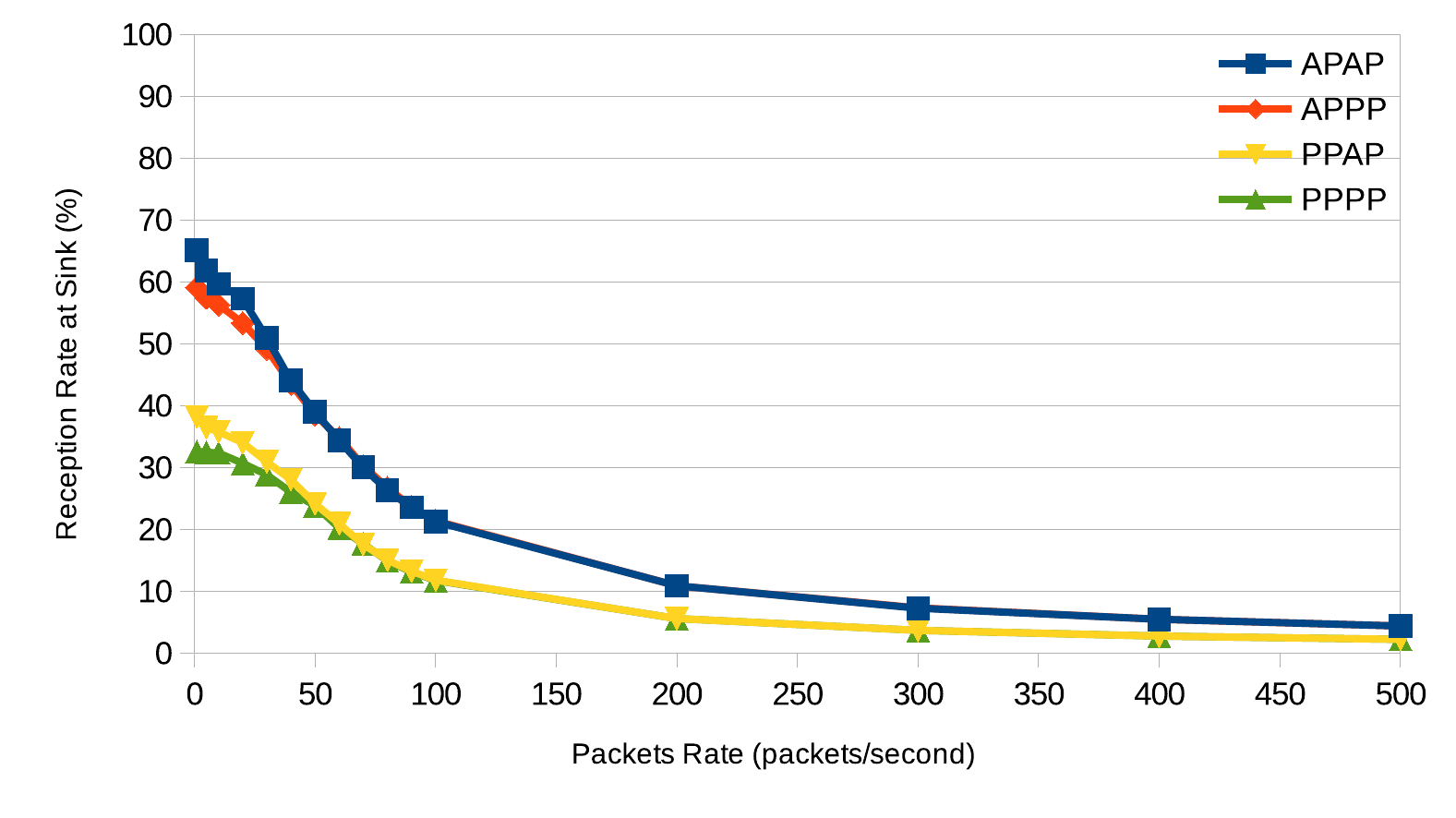, width=2.5in}
\caption{Reception Rate of Multi-Paths based strategies}
\label{Fig2}
\end{figure}

In \emph{Multi-Paths based strategies} (Figure \ref{Fig2}), the reception rate decreases fast with the increase of the transmission rate. APAP and APPP have better reception rate than PPAP and PPPP. The reason is that APAP and APPP strategies allow source nodes to send packets to both their parents to increase (in particular for sources who are farthest from the sink) the probability of their packets reception. Moreover, APAP is better than APPP and PPAP is better than PPPP, because in APAP and PPAP nodes have more chances to see their packets forwarded by their their parents than in APPP and PPPP, respectively.

\begin{figure}[!t]
\centering
\epsfig{file=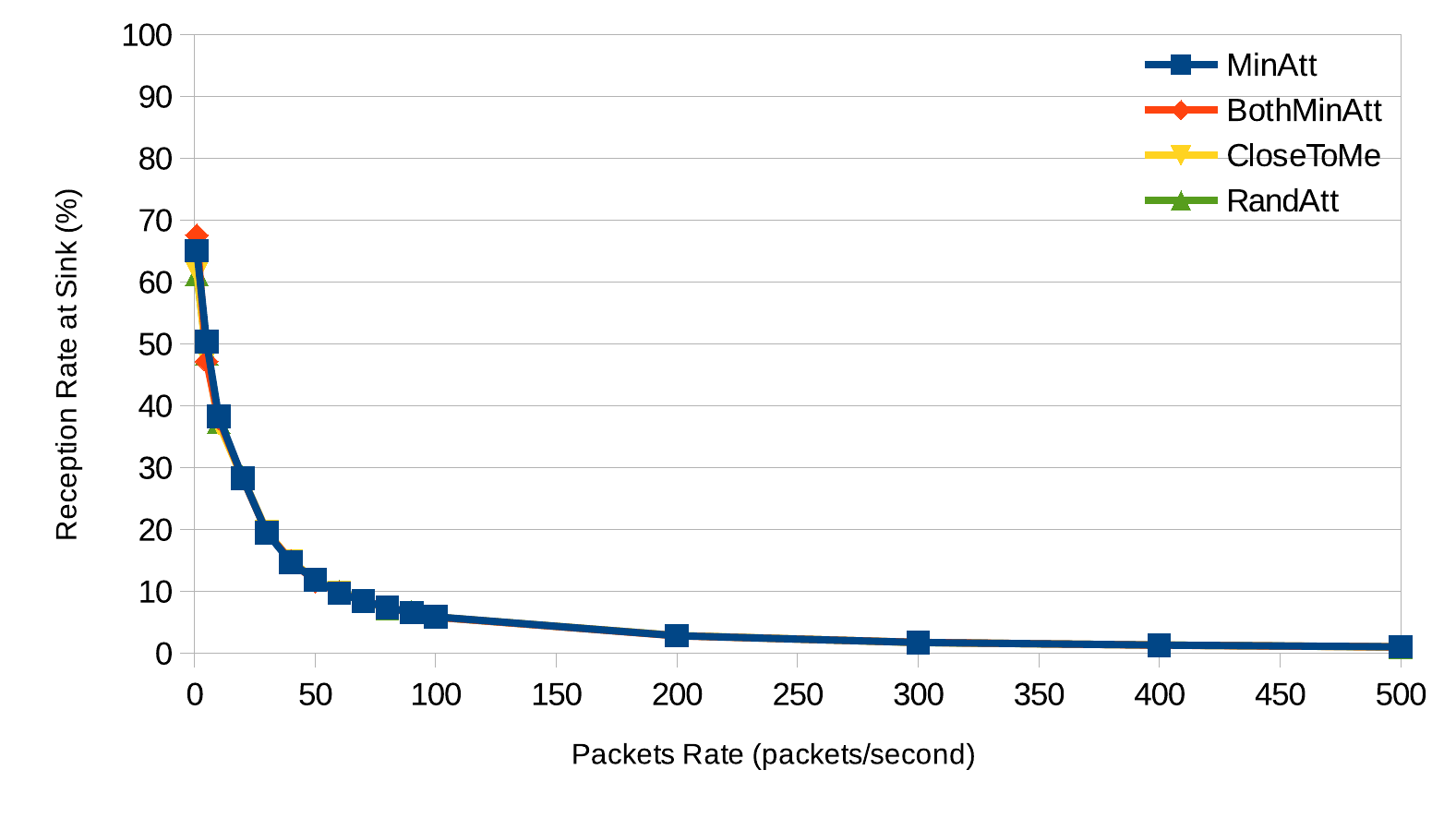, width=2.5in}
\caption{Reception Rate of Attenuation based strategies}
\label{Fig3}
\end{figure}

\emph{Attenuation-based strategies} (Figure \ref{Fig3}) have almost the same reception rate tendency and it decreases fast when transmission rate is superior to 50 packets par second. The reason is that the Request/Reply mechanism is inefficient and heavy in the intermittent environment of WBAN. The missing Reply for corresponding Request prevent sources from sending packets to next hop(s). Also, these strategies have important packets loss due to the over-buff of the waiting queues.
  
\begin{figure}[!t]
\centering
\epsfig{file=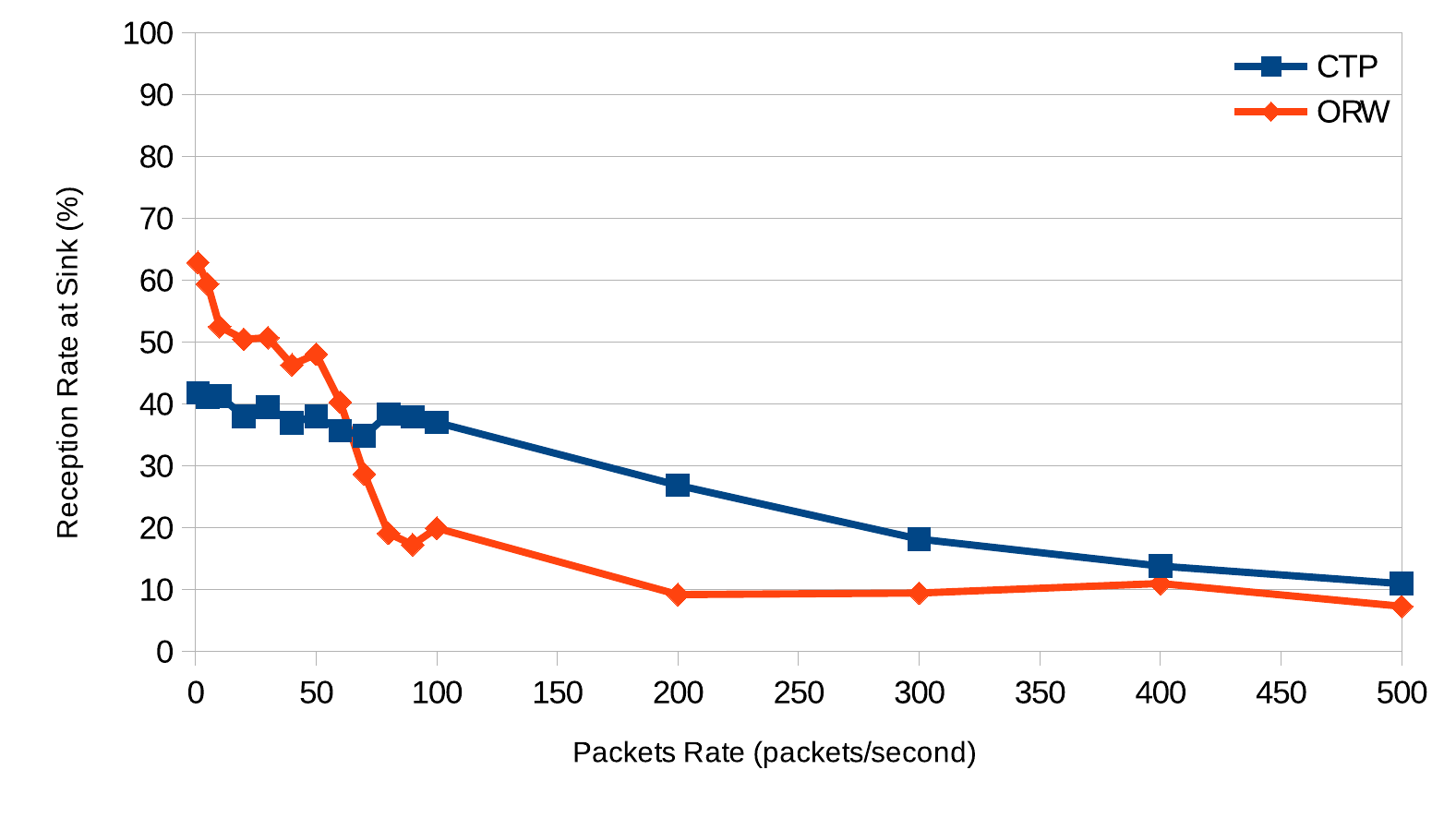, width=2.5in}
\caption{Reception Rate of Dynamic Path based strategies}
\label{Fig4}
\end{figure}

\emph{Dynamic Path strategies} simulations are presented in Figure \ref{Fig4}. ORW is better than CTP for transmission rates inferior to 75 packets par second. After that, ORW decrease faster that CTP. The reason is that CTP dynamically chooses the best next hop, but the in-band BEACON message affected the reception rate of useful packets. ORW chooses a set of next hops which increase the reception rate obviously when the transmission rate is not high enough. Furthermore, the increase of the transmission rate causes more and more collisions which yields to a fast decrease of the reception rate.

\begin{figure}[!t]
\centering
\epsfig{file=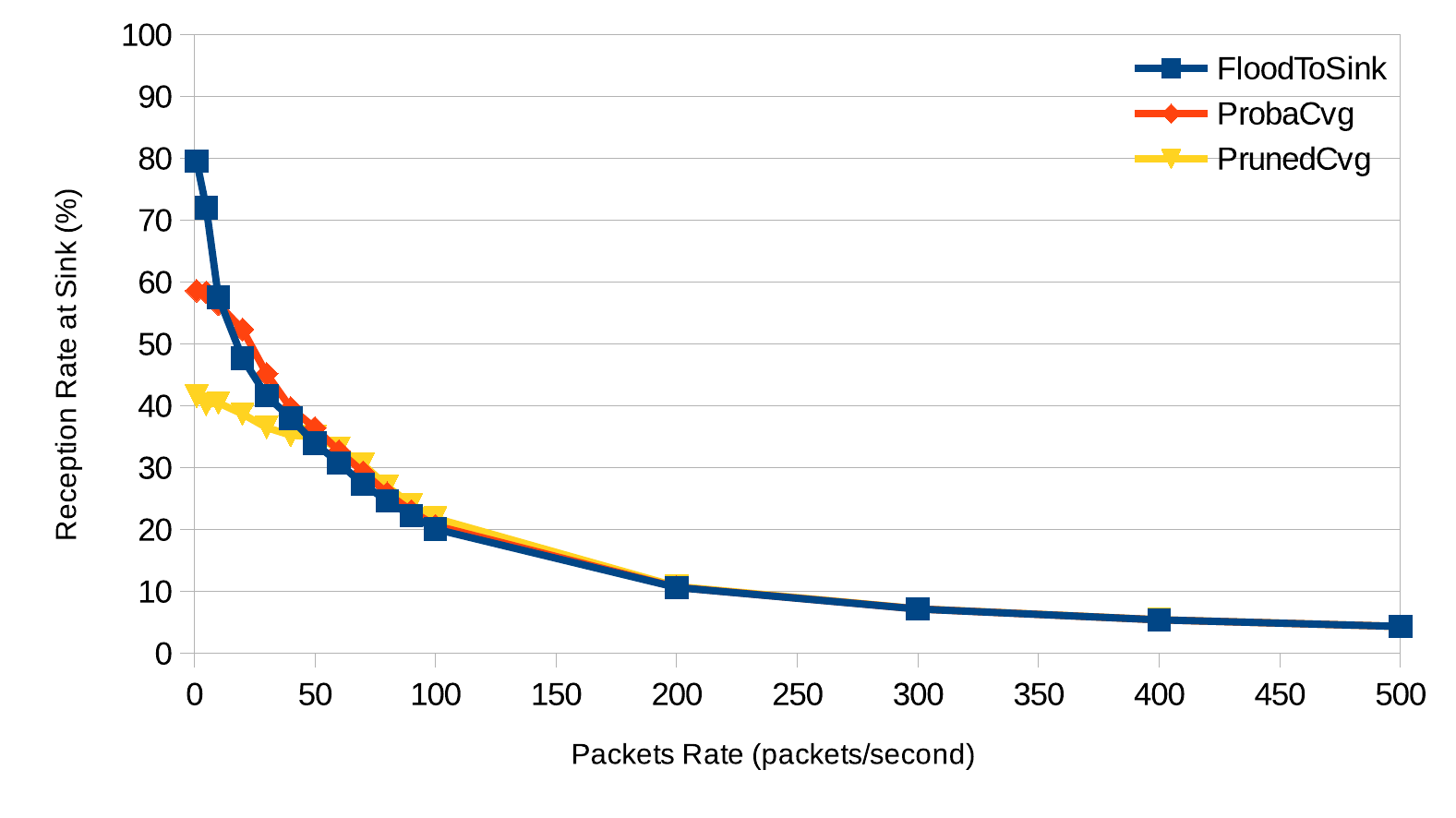, width=2.5in}
\caption{Reception Rate of Gossip-based strategies}
\label{Fig5}
\end{figure}

In the set of \emph{Gossip-based strategies} (Figure \ref{Fig5}), FloodToSink shows a good reception rate at the beginning, but from the same reason as ORW, the reception rate decreases fast. ProbaCvg and PrunedCvg like the limitation version of the FloodToSink show a worse and worst result respectively at the beginning, but they decrease not as fast as FloodToSink due to the fewer packets flooding into the network in ProbaCvg and PrunedCvg.

The percentage of delivered messages at sink is relatively low and the reception rate decreases rapidly with the increase of the traffic rate.

The reason for the low reception rate is the intermittent network connection due to the important attenuation of the channel-mobility model. Reaching of the limitation of network throughput is the reason for the rapid decreases of reception rate with the growing of data rate. The network reaches its limitations due to more and more collisions, interferences, packet and over-buff errors. 

%For CTP and ORW, we take the means of 10 simulation RUNs as the results, but there are still some irregular points in the fugues, where with higher Packets Rate, the Reception Rate increase instead of decrease. The reason is that CTP and ORW are strategies who update path during the packets transmission, so the random elements in the simulations, like the backoff of collisions, the attenuation and the interference between packets, will have a bigger influence to the path choosing and to the reception rate than other strategies. So for every different simulation RUN, the results of CTP and ORW have more variance than others, especially when two point has a small difference of packets rate (in other words, when two point are close to each other). This situation show that CTP and ORW are less stable for each RUN and have more varying than other strategies.
 
\subsubsection{Total order Analysis}
In the section we analyse the ability of the studied strategies to deliver packets at the sink respecting the total order property: packets from the same source should be received at the sink in the same order as they have been sent. 
From our simulations none of the 13th strategies has a zero wrong-sequence packets. See Figures \ref{Fig7}, \ref{Fig8} and \ref{Fig9} for a zoom on the packet rates in between 1 and 50 per second. In these figures the green curve corresponds to the percentage of total ordered messages while yellow curves corresponds to the percentage of delivered messages. 
The main reason for this wrong-sequence delivery is the fact that all strategies create virtual multi-paths during packet transmissions.

\emph{
We performed additional simulations concerning traditional mechanisms such as retransmissions and sequences. Due to lack of space, we only resume our conclusions.
The reception rate when using the ACK-based retransmission is worse than the case of no retransmission is used. When using sequencer mechanisms at the sink level we obtain a end-to-end delay very important or we have to sacrifice the rate reception.
The analysis performed in the previous sections and additional simulations focusing the traditional mechanisms (e.g. retransmissions and sequencers) conducted us to propose a new reliability mechanism based on a new channel model and a new convergecast strategy that outperform existing one.
}
\section{New model}
\label{model}
%In this section, we propose a new WBAN model. Similar to Time-Varying Graphs (TVG)\cite{casteigts2012time} used in the dynamic networks studying, the new WBAN model bases on no more the Determinist Time-Varying Graphs but a Probabilist Posture-Varying Graphs (PPVG), where we focus on the connexion existing probability in different postures.

%For that, we need to evaluate our channel-mobility model, in other words, the channel attenuation. Attention of different links (channel between each two nodes) are random values from different normal distributions known already means and standard deviations from the metric of attenuation. 
From the mean and the standard deviation for each link in \cite{naganawa2015simulation}, we can thus calculate the Cumulative Distribution Function (CDF) of the random attenuation: $F(X) = P[x<X]$. 
$F(X)$ is the probability of the random event: $x$ is smaller than $X$, where x is the random attenuation variable and X is a threshold. According to our channel-mobility model, links appear and disappear from time to time, because of the channel attenuation. A connexion exists if the transmission power minus the attenuation is bigger than the sensibility and doesn't exist if not.
%:
%
%Ptx - Att $>$ Sensibility: connexion exists.
%
%Ptx - Att $<=$ Sensibility: connexion doesn't exist.

So if $X$ is the maximum acceptable attenuation in a link for transmission, $F(X)$ is the probability that the random attenuation is smaller than the maximum acceptable attenuation. In order words the connexion exists and the transmission of this time could be successful. $1/F(X)$ is thus the Expected Transmission Count (ETX) in this link for a transmission.

In our case, we use $-60dBm$ as the initial transmission power, $-100dBm$ the sensibility. It follows that the maximum acceptable attenuation is $(-60dBm - (-100dBm)) = 40dB$. If $X = 40dB$, then $F(40)$ is the probability for a successful transmission and $1/F(40)$ is the $ETX$ of the transmission in this link. Since we have all the means and standard deviations so we can compute all the $ETX$ for every link.

%\subsection{Posture Division Topologies}

\begin{figure}[!t]
\centering
\epsfig{file=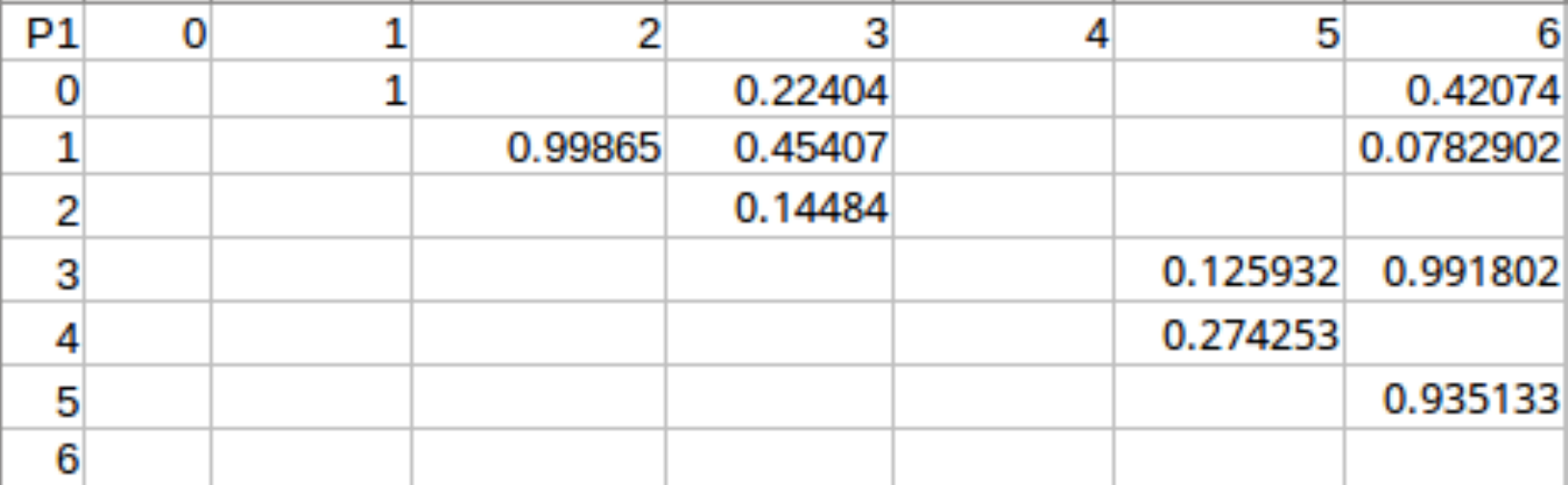, width=2.5in}

\centering
\epsfig{file=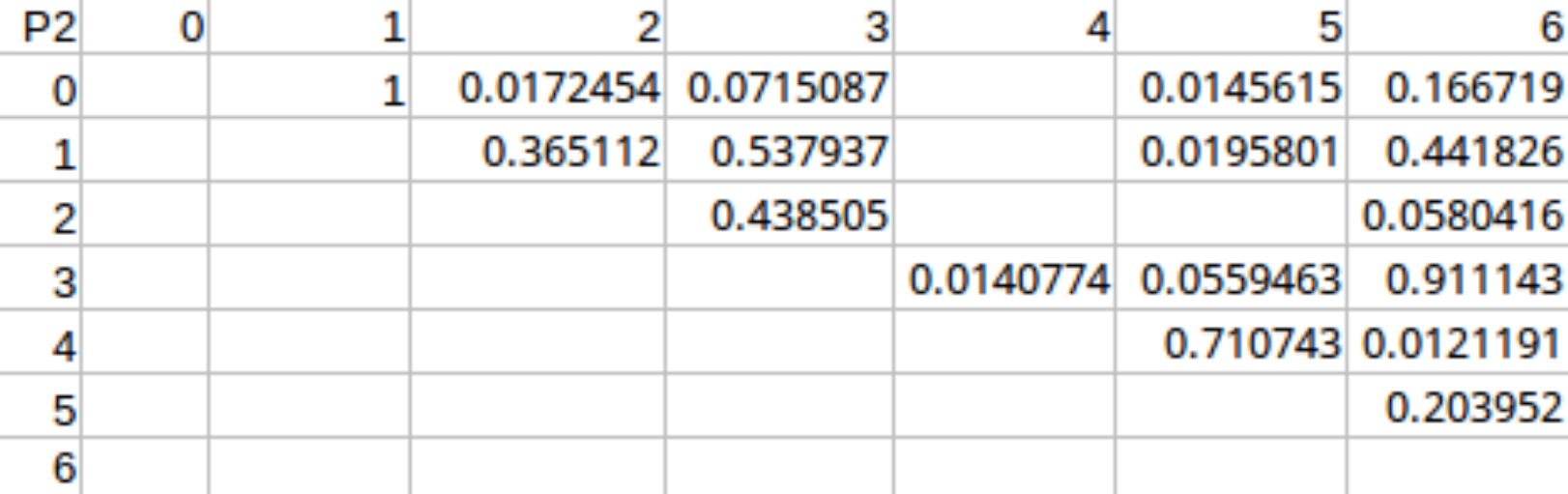, width=2.5in}

\centering
\epsfig{file=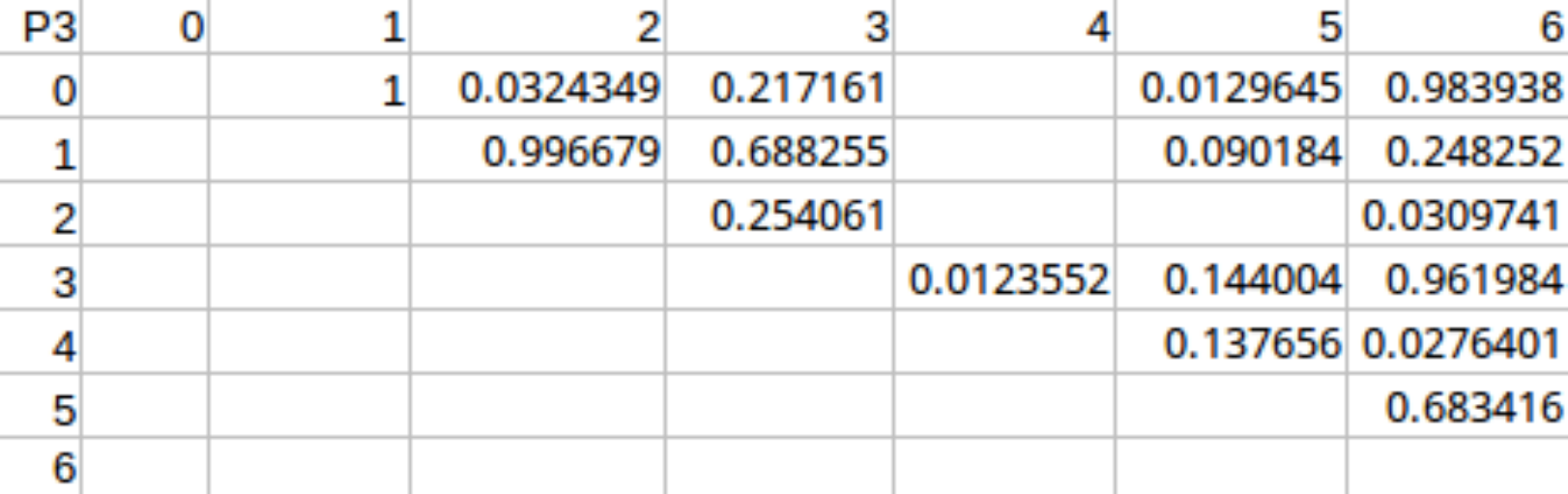, width=2.5in}

\centering
\epsfig{file=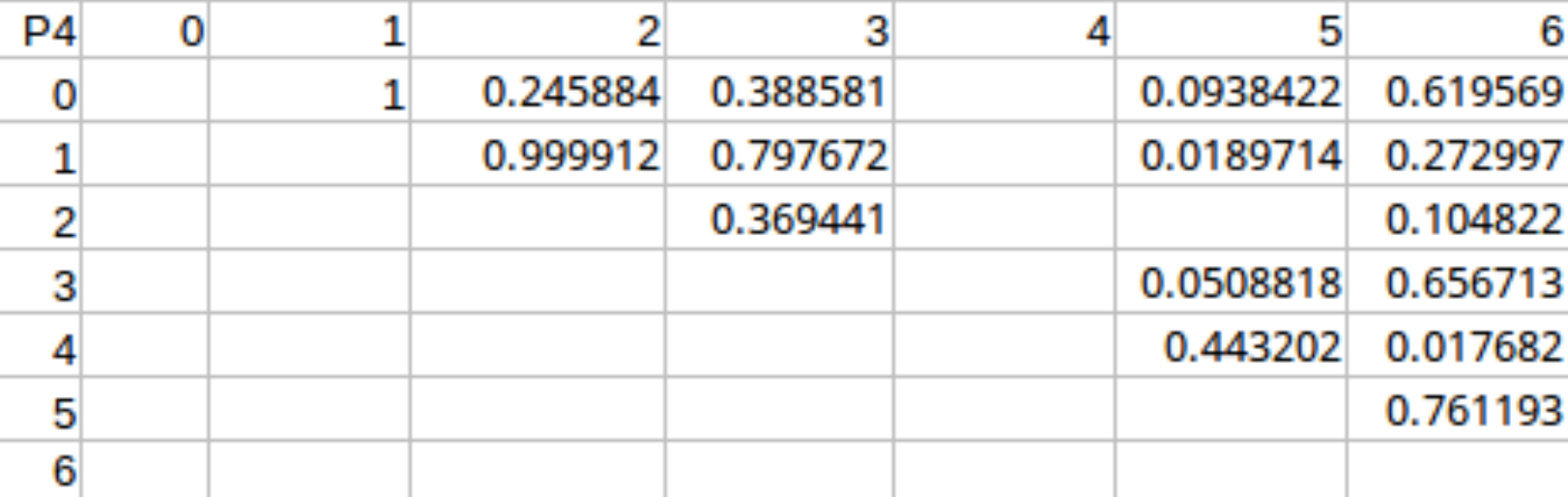, width=2.5in}

\centering
\epsfig{file=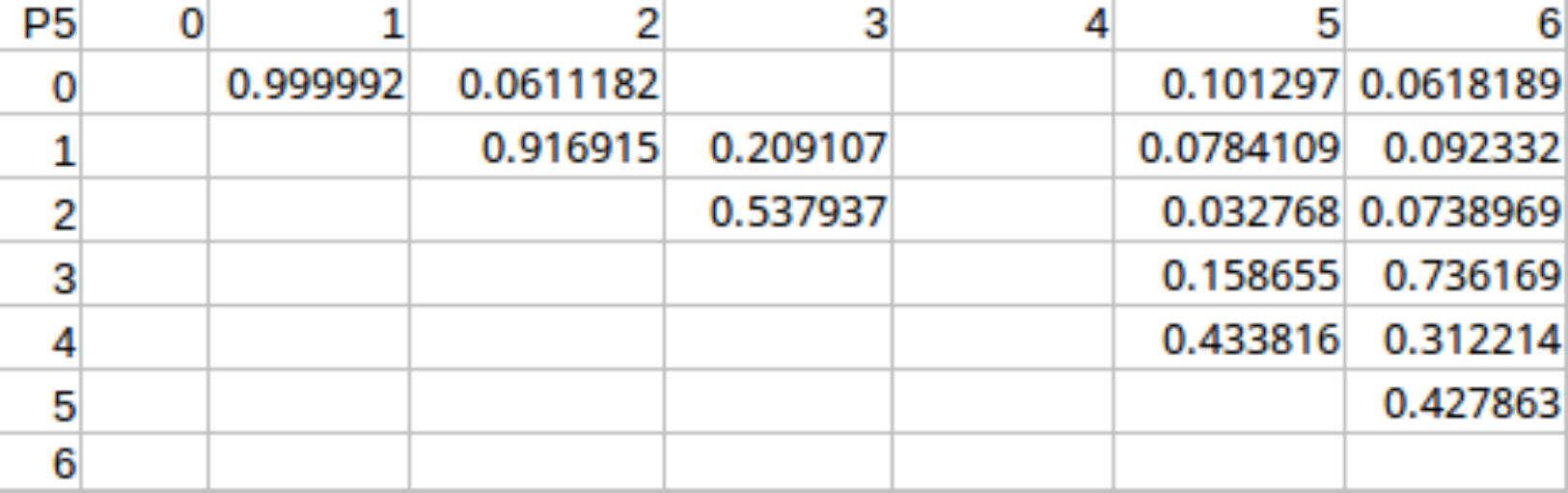, width=2.5in}

\centering
\epsfig{file=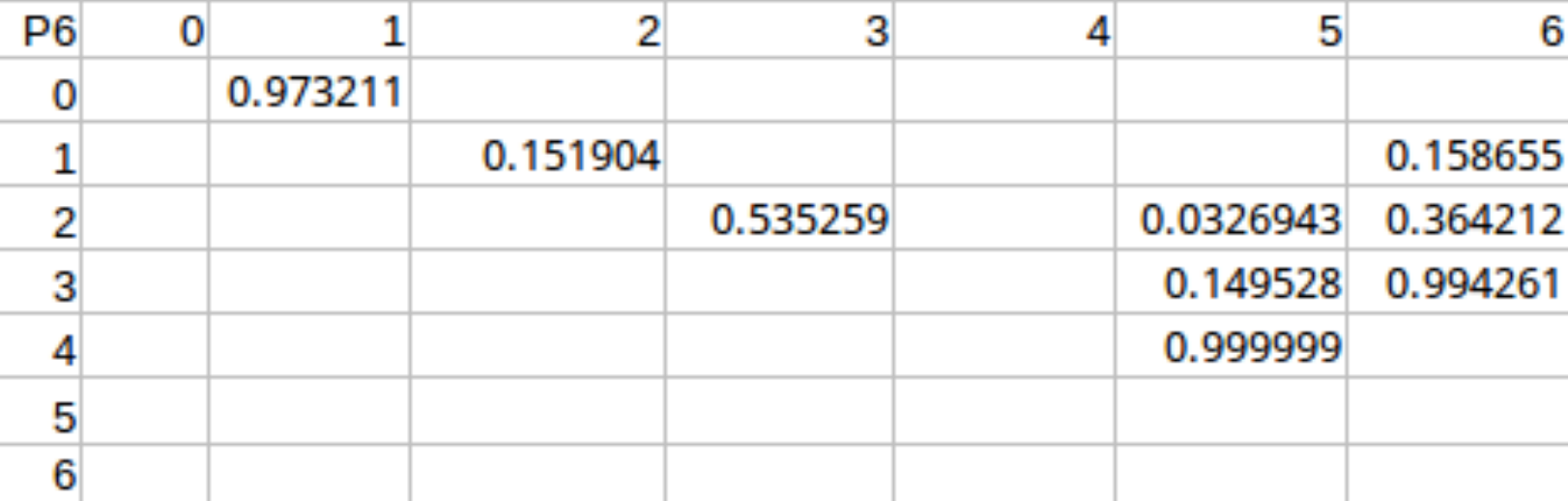, width=2.5in}

\centering
\epsfig{file=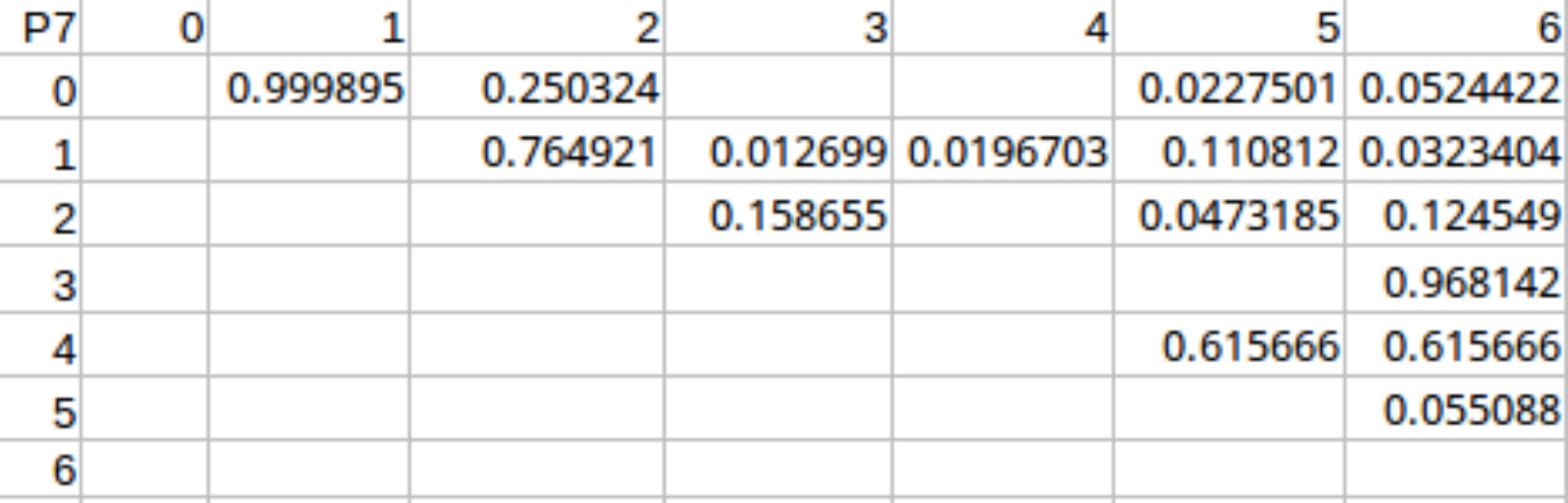, width=2.5in}
\caption{Links-existing probability for each link in 7 Postures}
\label{Fig6}
\end{figure}

%According to the theory of No-ACK Retransmission, as long as we know the sender, the receiver, the current mobile posture and model channel-mobility information, we can get the probability of every link between the sender and the receiver in each different posture.
Figure \ref{Fig6} show all the connexions probability for each link in different postures. In the sequel we consider only  the probabilities greater than 0.01. 
%From all these links probability, we get the Probabilist Posture-Varying Graphs of WBAN.

\section{New Reliability Mechanism}
\label{mechanism}

In the following we propose a No-ACK Retransmission Mechanism based on our channel-mobility model. The idea is that each node continuously transmits the same packet expected retransmission count ($ETX$) times without waiting for ACK. The $ETX$ value is according to the model described in the previous section. After sending a packet $ETX$ times, nodes continue to send next packet with this same mechanism.

%Traditionally in the retransmission based on ACK, the sender continues to re-send packets until it receives the corresponding ACK replied by receiver or the max retransmission number is reached. However, this kind of mechanism usually needs reliable connexions among 1-Hop neighbors, if not, the missing ACK will lead to unnecessary retransmissions. We implemented ACK-based retransmission in CTP and ORW and obseved that the reception rate is worse than the case when no ACK is used (see Figure \ref{Figctporw}). 
%Due to the lack of space the simulation results for this case are not reported here.
%Without the information of ACK, we need to know, for each time and each connexion between two nodes: 

%According to the sender, the receiver and the ETX for the link between sender and receiver, each time when sender has packets to send, it will send each packet ETX times without waiting for any reply (ACK). So we make sure, theoretically, that packets could be received after ETX times retransmissions, the treating time for each packet is relatively small and the amount of useless packets sent into the network are also small. The receiver will treat the copies of packet only one time to reduce unnecessary retransmission.

%\subsection{Simulation Result}

\begin{figure}[!t]
\centering
\epsfig{file=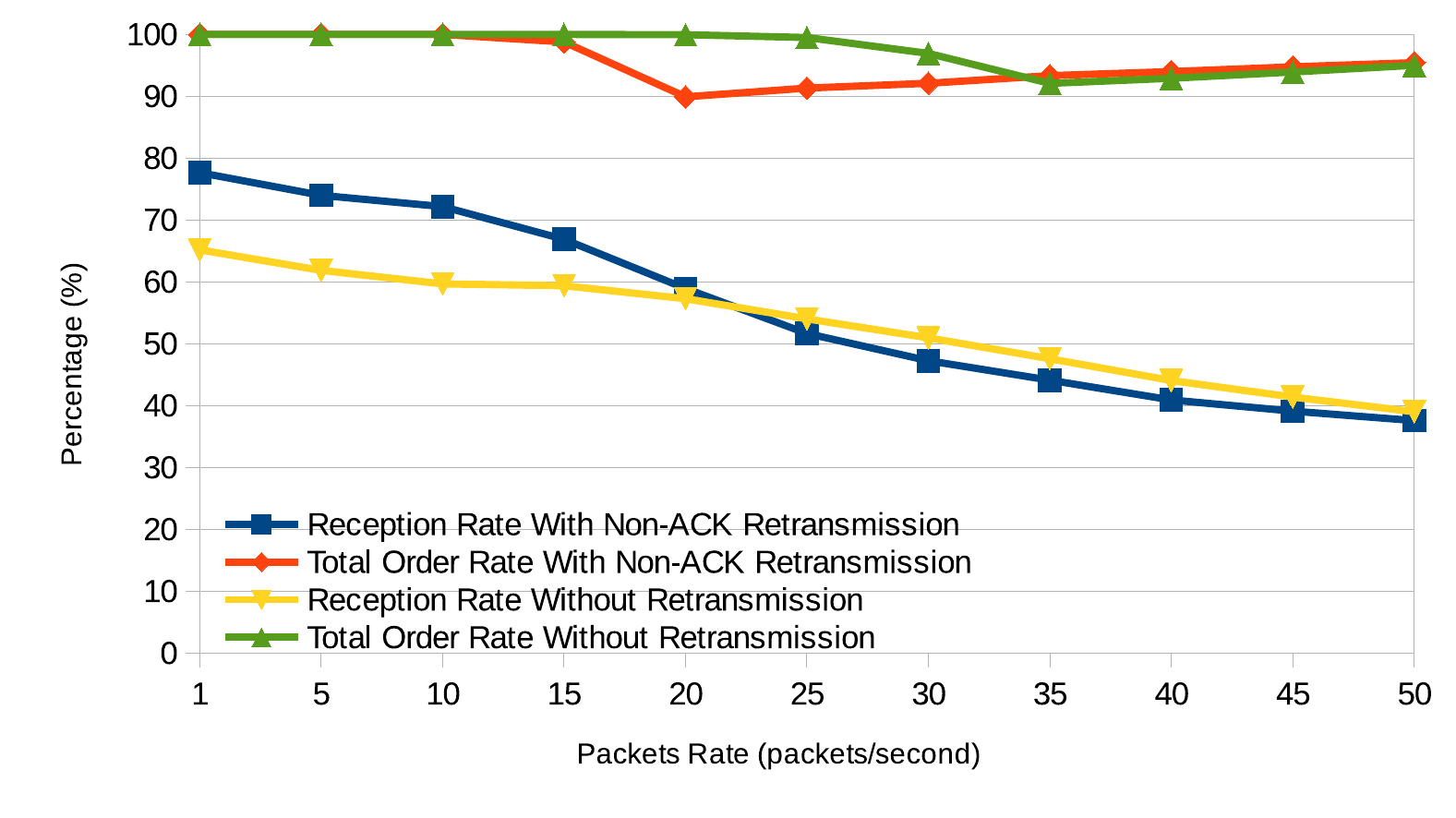, width=2.5in}
%\caption{Multi-Paths based strategies}

\centering
\epsfig{file=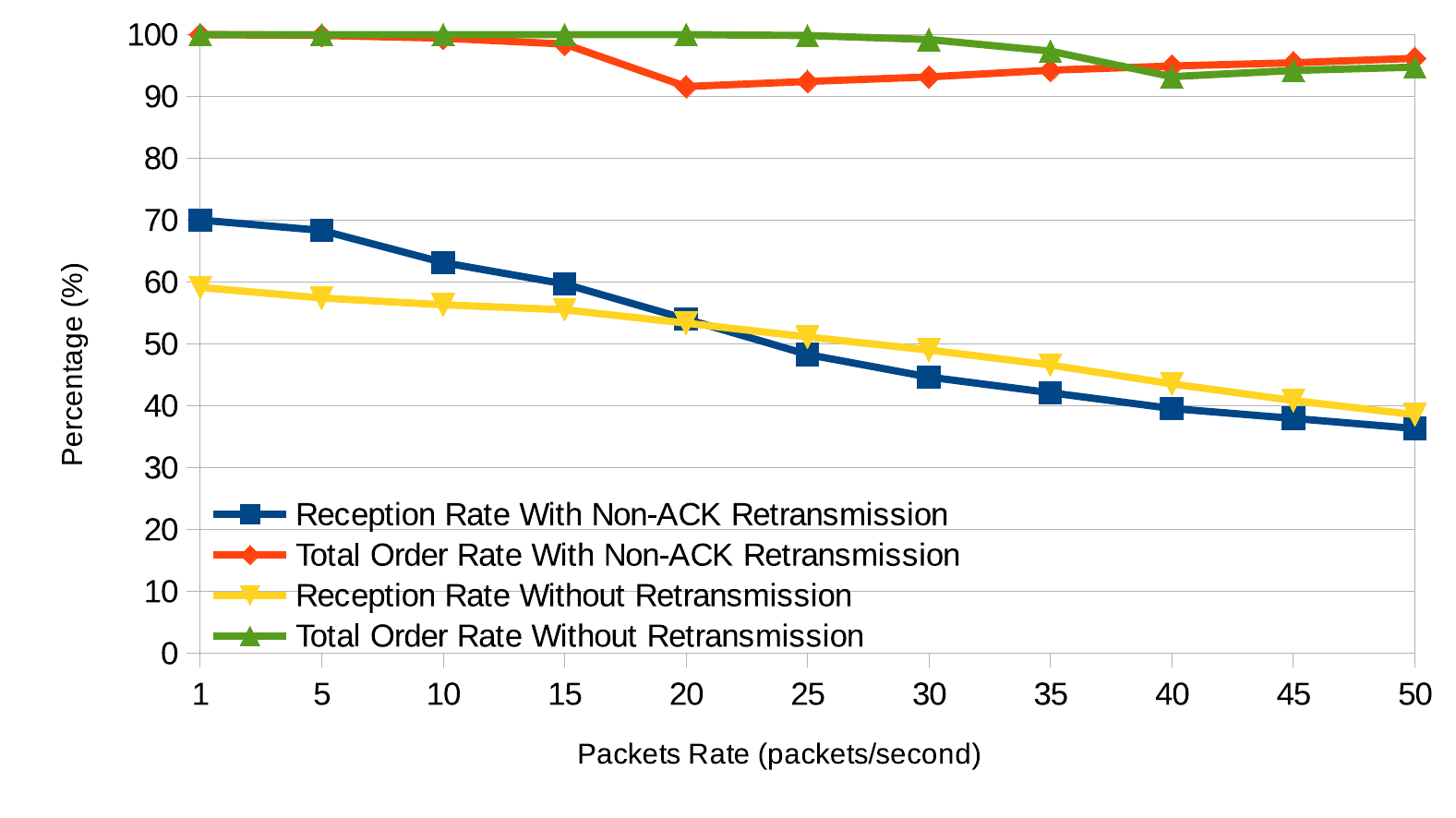, width=2.5in}
%\caption{Multi-Paths based strategies}
%\label{ExecutionTime}

\centering
\epsfig{file=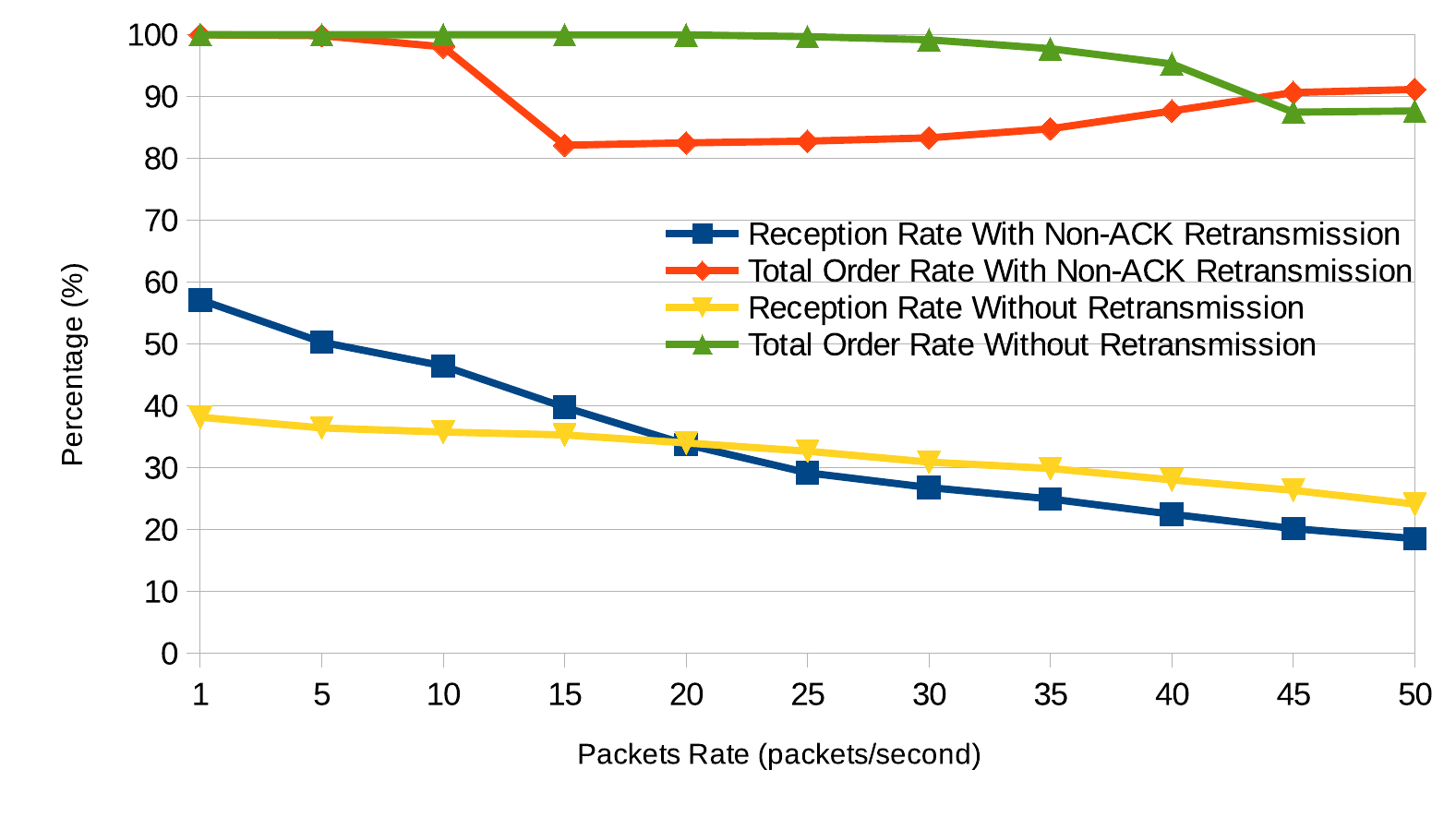, width=2.5in}
%\caption{Multi-Paths based strategies}
%\label{ExecutionTime}

\centering
\epsfig{file=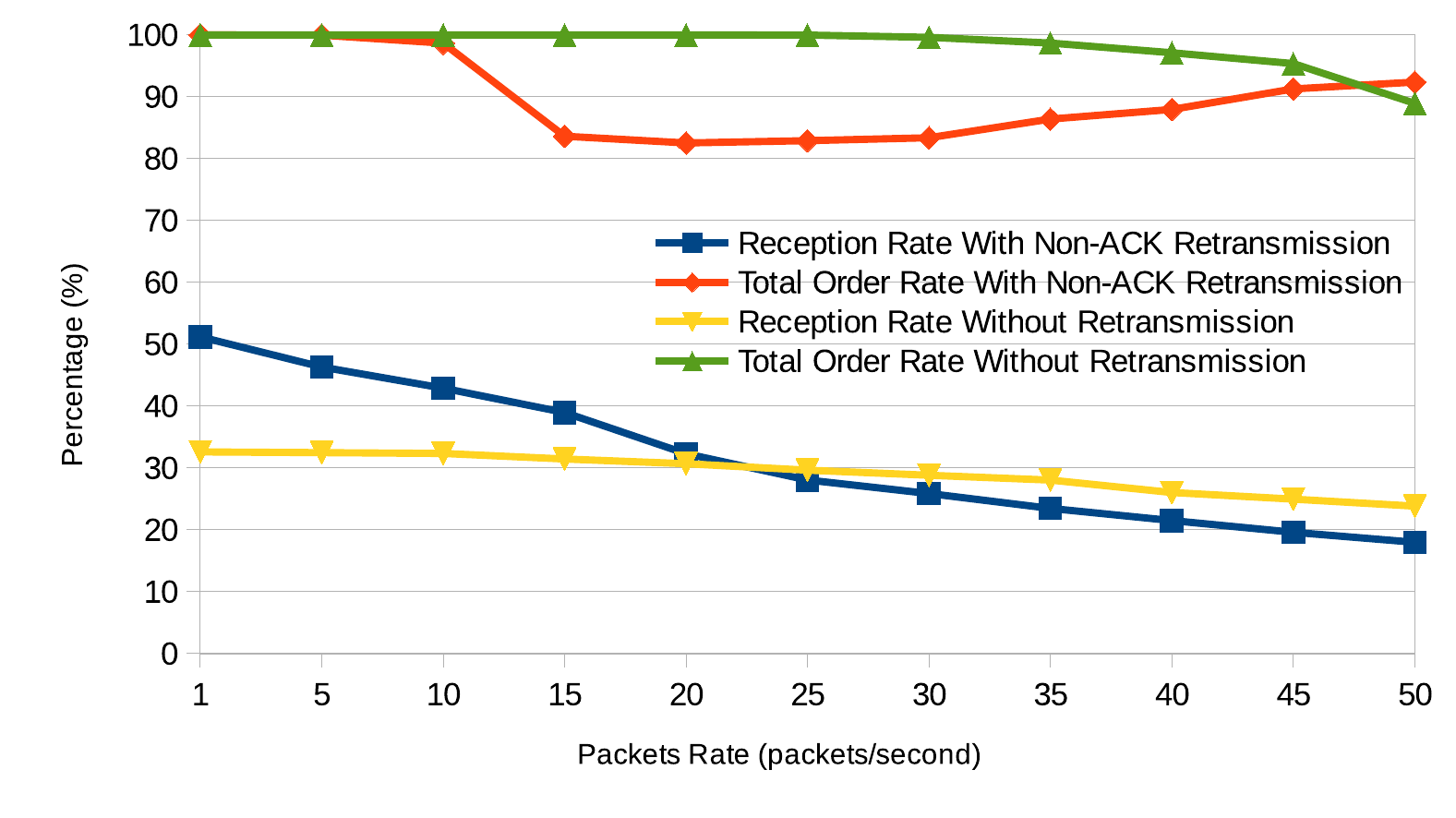, width=2.5in}
\caption{Comparison of using No-ACK Retransmission or not in term of Reception Rate and Total Order Rate for Multi-Paths based Strategies}
\label{Fig7}

\end{figure}

\begin{figure}[!t]
\centering
\epsfig{file=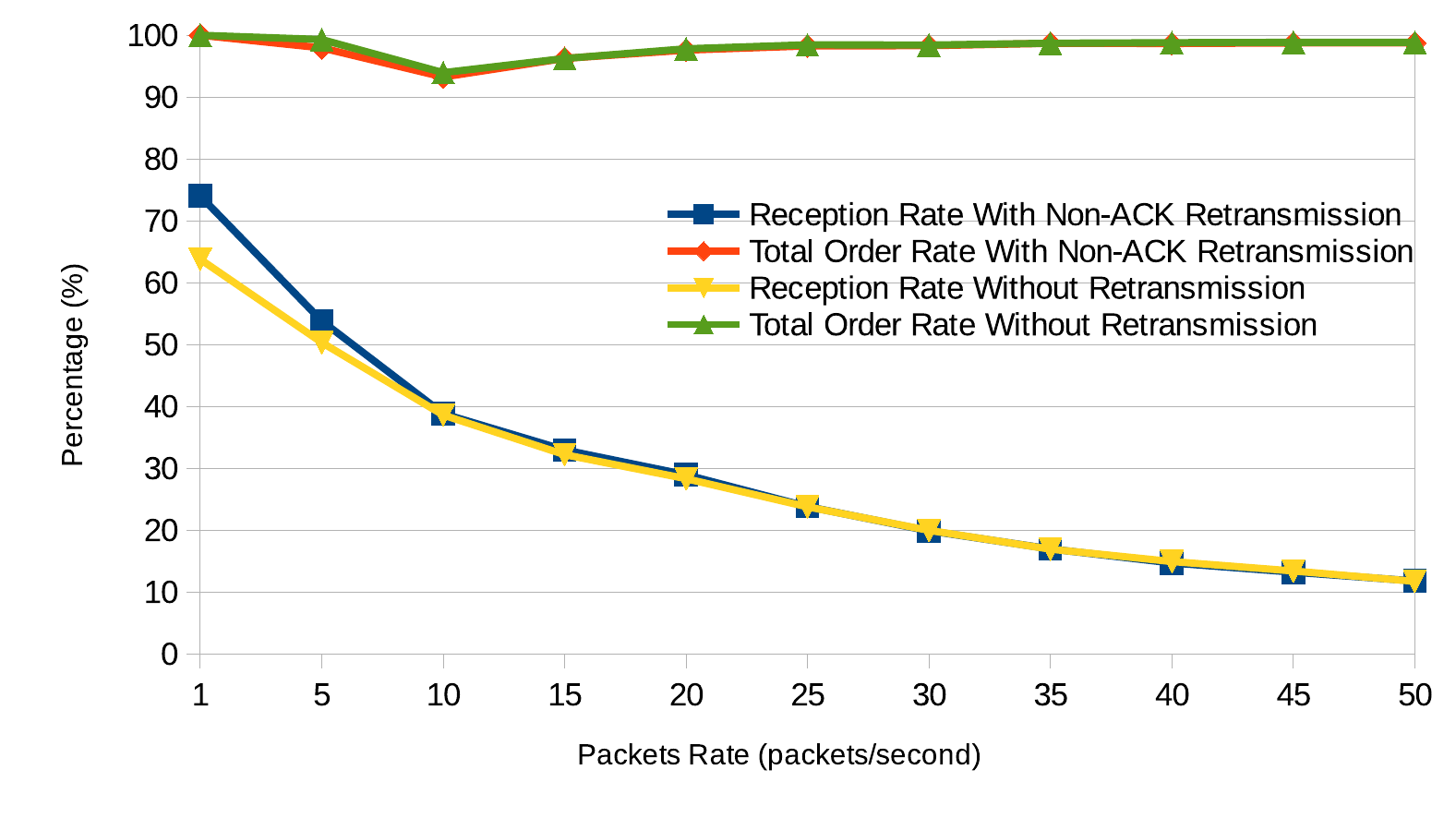, width=2.5in}
%\caption{Multi-Paths based strategies}
%\label{ExecutionTime}

\centering
\epsfig{file=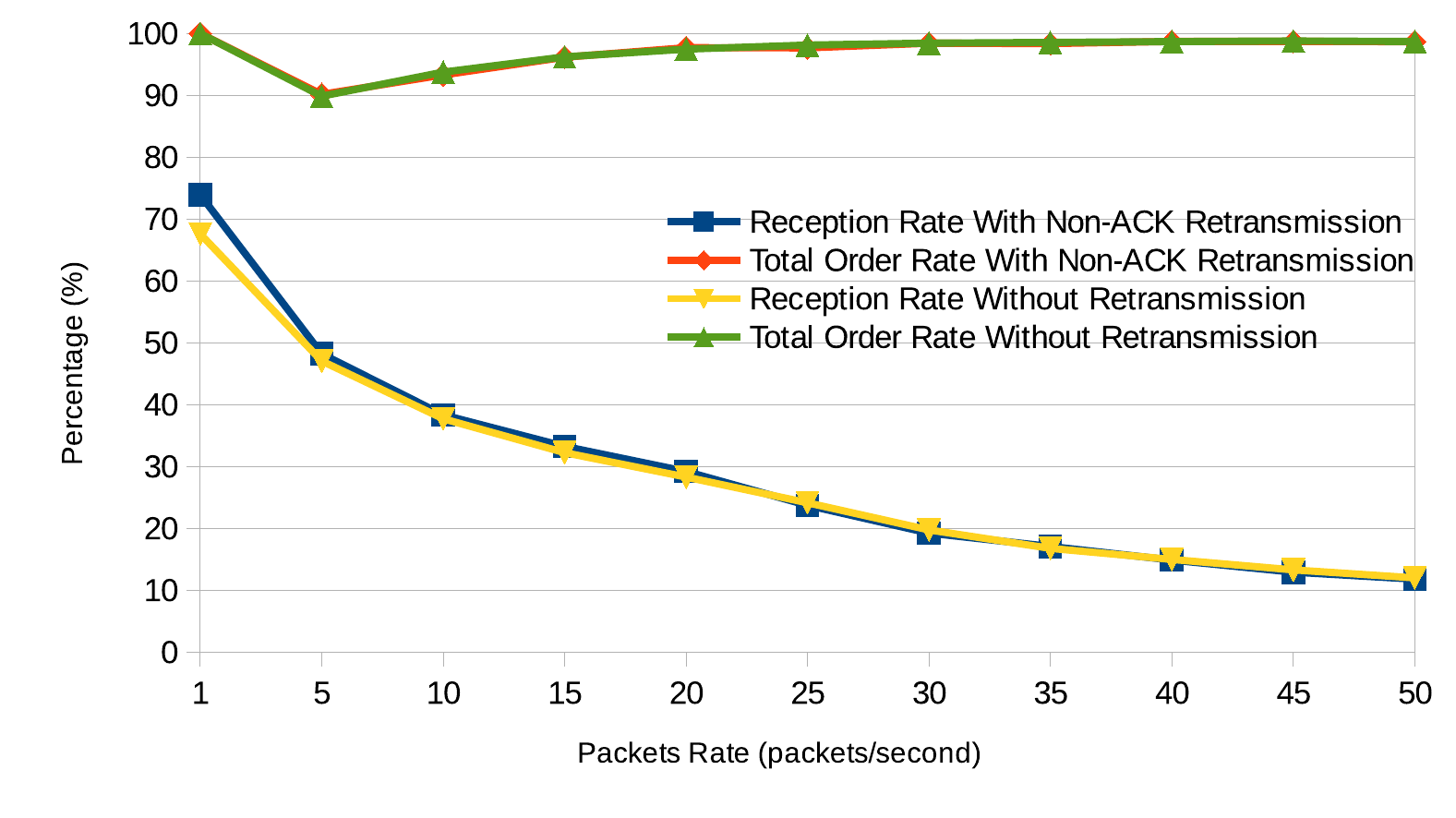, width=2.5in}
%\caption{Multi-Paths based strategies}
%\label{ExecutionTime}

\centering
\epsfig{file=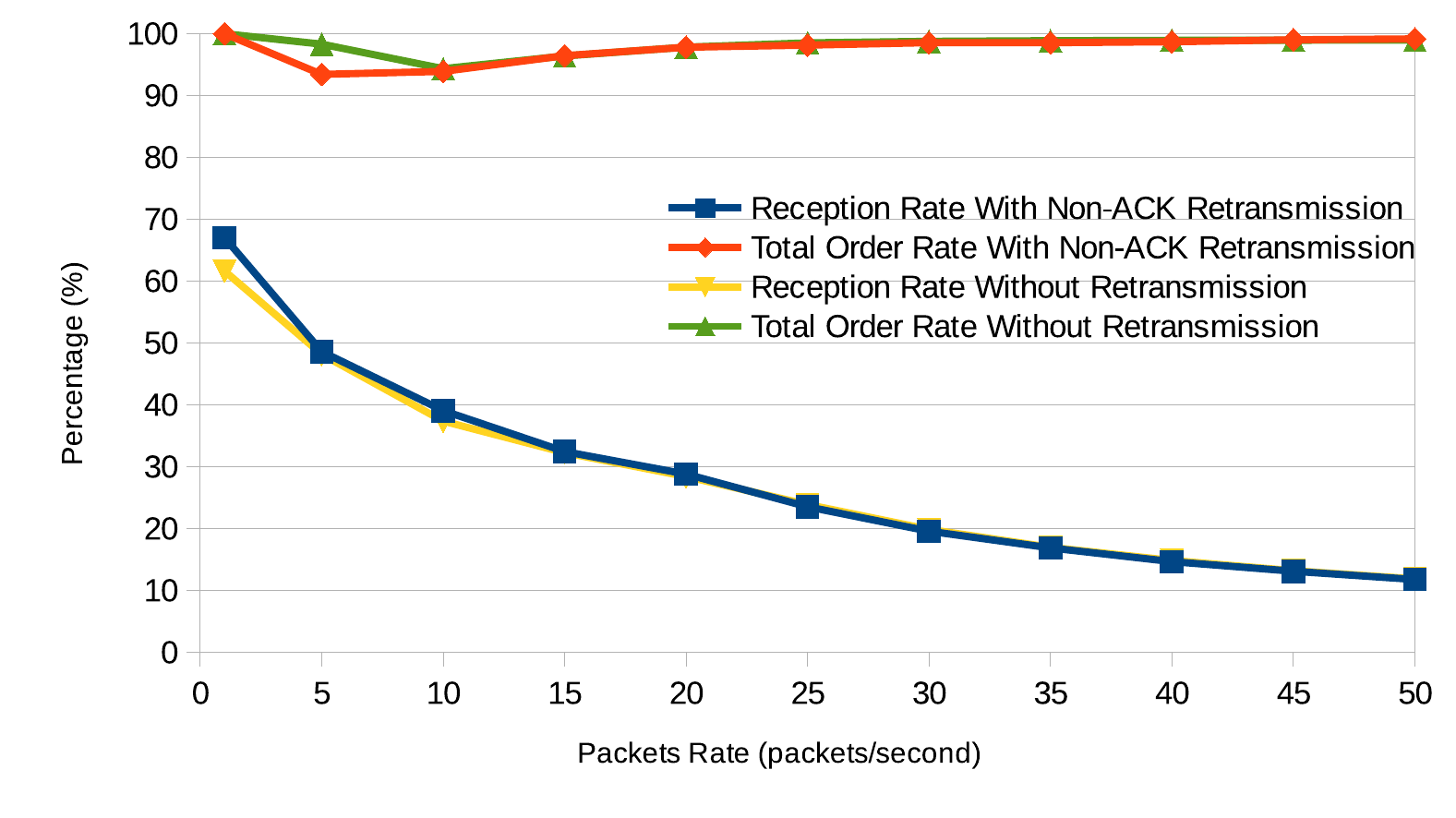, width=2.5in}
%\caption{Multi-Paths based strategies}
%\label{ExecutionTime}

\centering
\epsfig{file=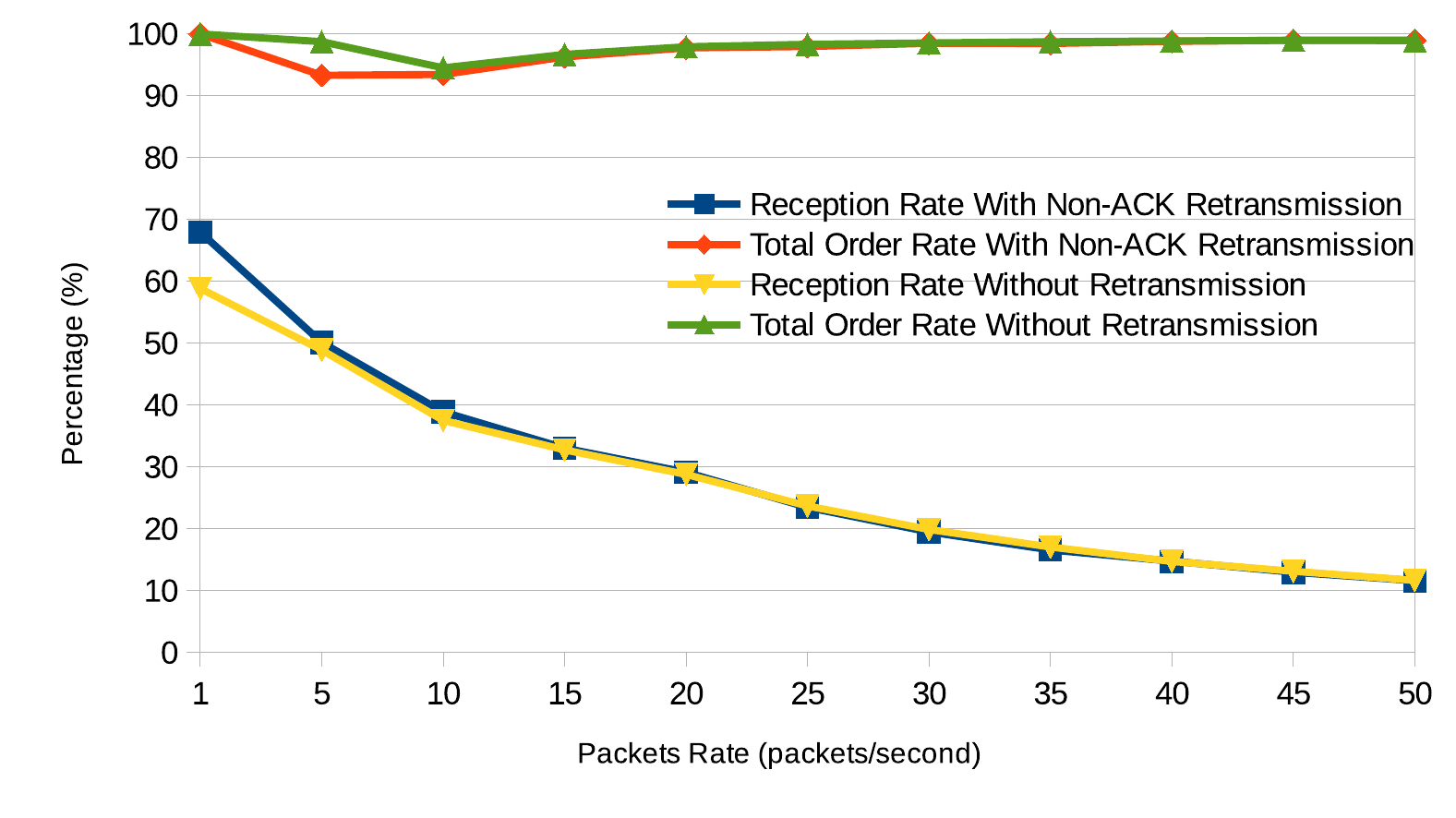, width=2.5in}
\caption{Comparison of using No-ACK Retransmission or not in term of Reception Rate and Total Order Rate for Attenuation-based Strategies}
\label{Fig8}

\end{figure}

\begin{figure}[!t]
\centering
\epsfig{file=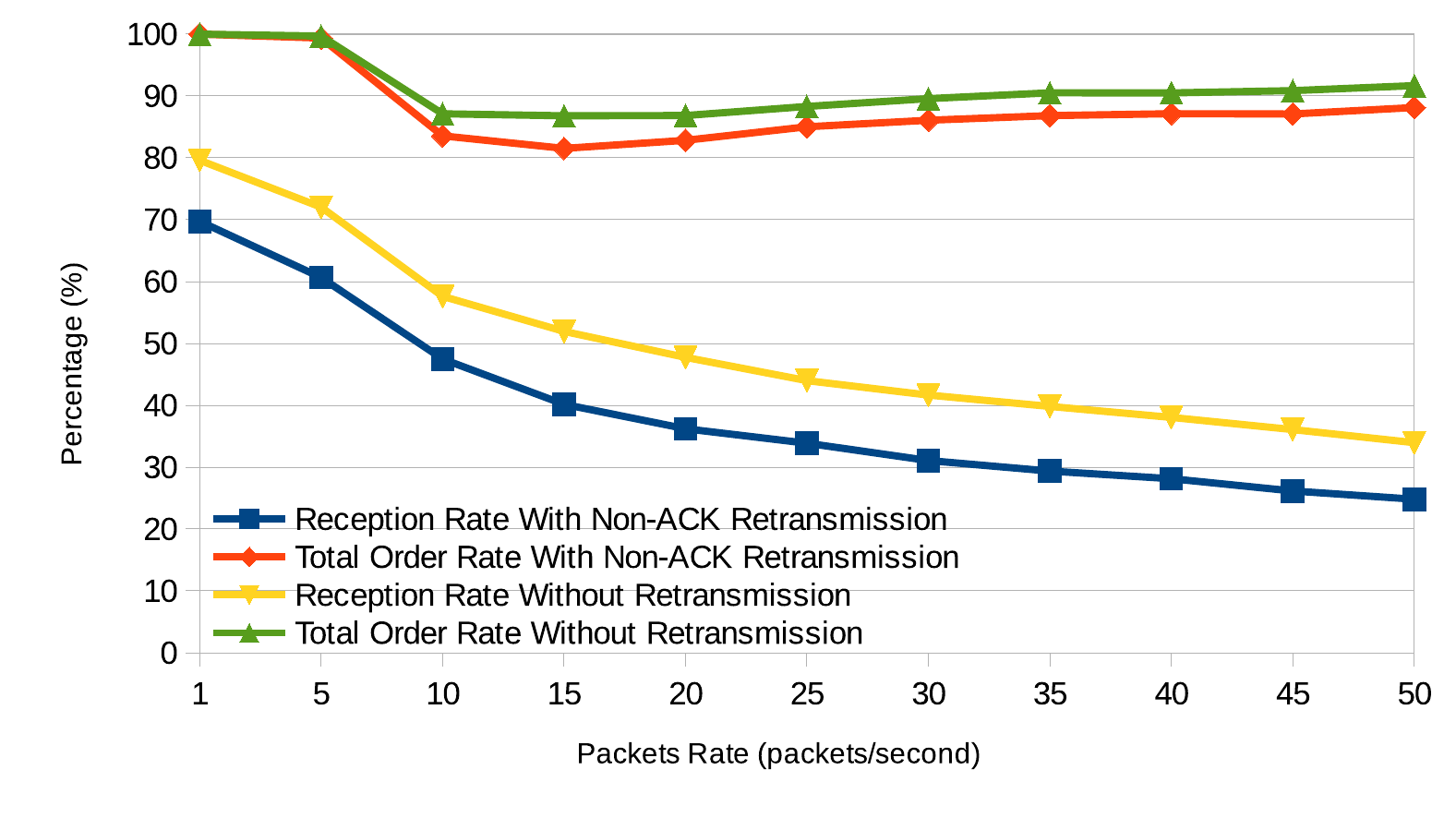, width=2.5in}
%\caption{Multi-Paths based strategies}
%\label{ExecutionTime}

\centering
\epsfig{file=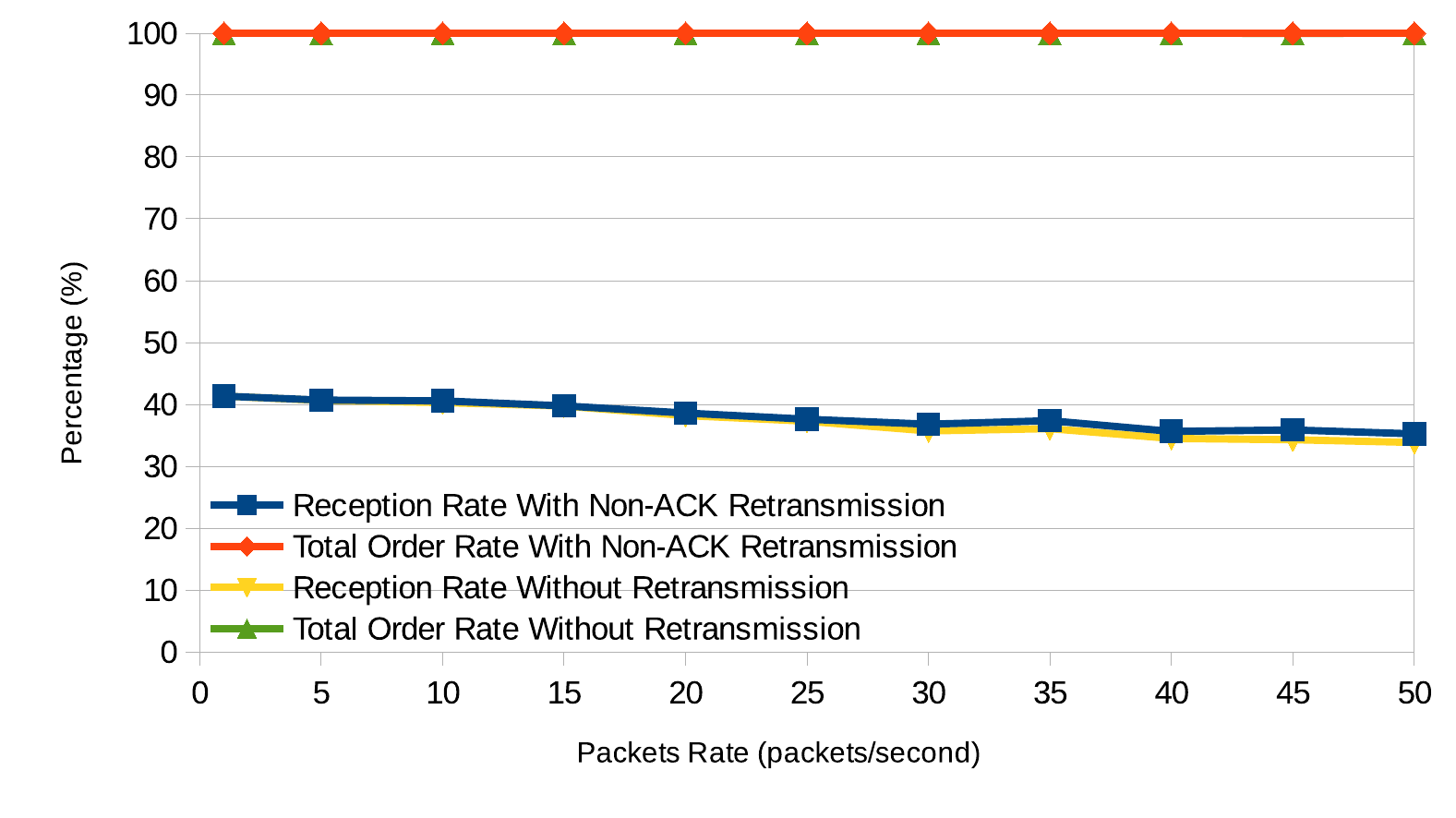, width=2.5in}
%\caption{Multi-Paths based strategies}
%\label{ExecutionTime}

\centering
\epsfig{file=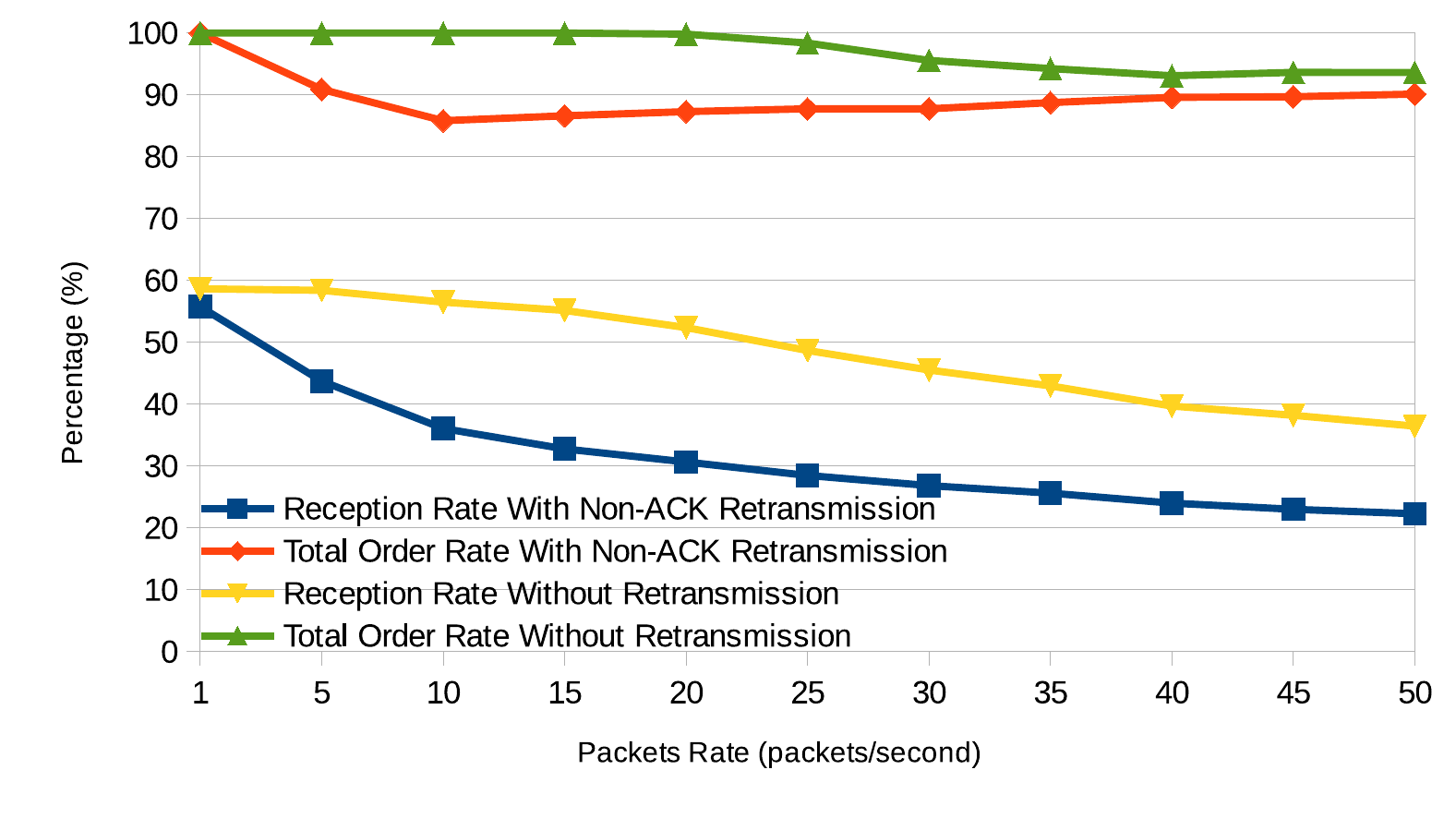, width=2.5in}
\caption{Comparison of using No-ACK Retransmission or not in term of Reception Rate and Total Order Rate for Gossip-based Strategies}
\label{Fig9}

\end{figure}

Figure \ref{Fig7}, Figure \ref{Fig8} and Figure \ref{Fig9} represent the comparative results between original strategies and strategies using our reliability mechanism for Multi-Paths based strategies, Attenuation-based strategies and Gossip-based strategies. Green and Red curves report the percentage of messages that respect the total order for the original strategies, respectively enhanced strategies. Yellow and Blue curves report the percentage of delivered messages for the original strategies, respectively enhanced strategies.

Our retransmission mechanism do help to improve the reliability, but with certain limitations: 

For Multi-Paths based strategies, our No-ACK mechanism improves the reception rate compared with original strategies up to a transmission rate of 20 packets par second. 
%But the decrease speed of strategies using our mechanism is faster than the original strategies.

For Attenuation-based strategies, our mechanism does not show a great improvement.  

For Gossip-based strategies our mechanism makes the reception rate lower than the originals. Note that ProbaCvg strategy will not be well affected by the retransmission mechanism because of its special particularity.

In terms of total order delivery our mechanism does not bring any significant improvement. It just brings forward the valley of the correct-sequence rate.
%, according to the formula we mentioned in the previous sections,
The reason is the augmentation of the transmission number compared with the original one.

\section{New convergecast strategy}
\label{strategy}
\begin{figure}[!t]
\centering
\epsfig{file=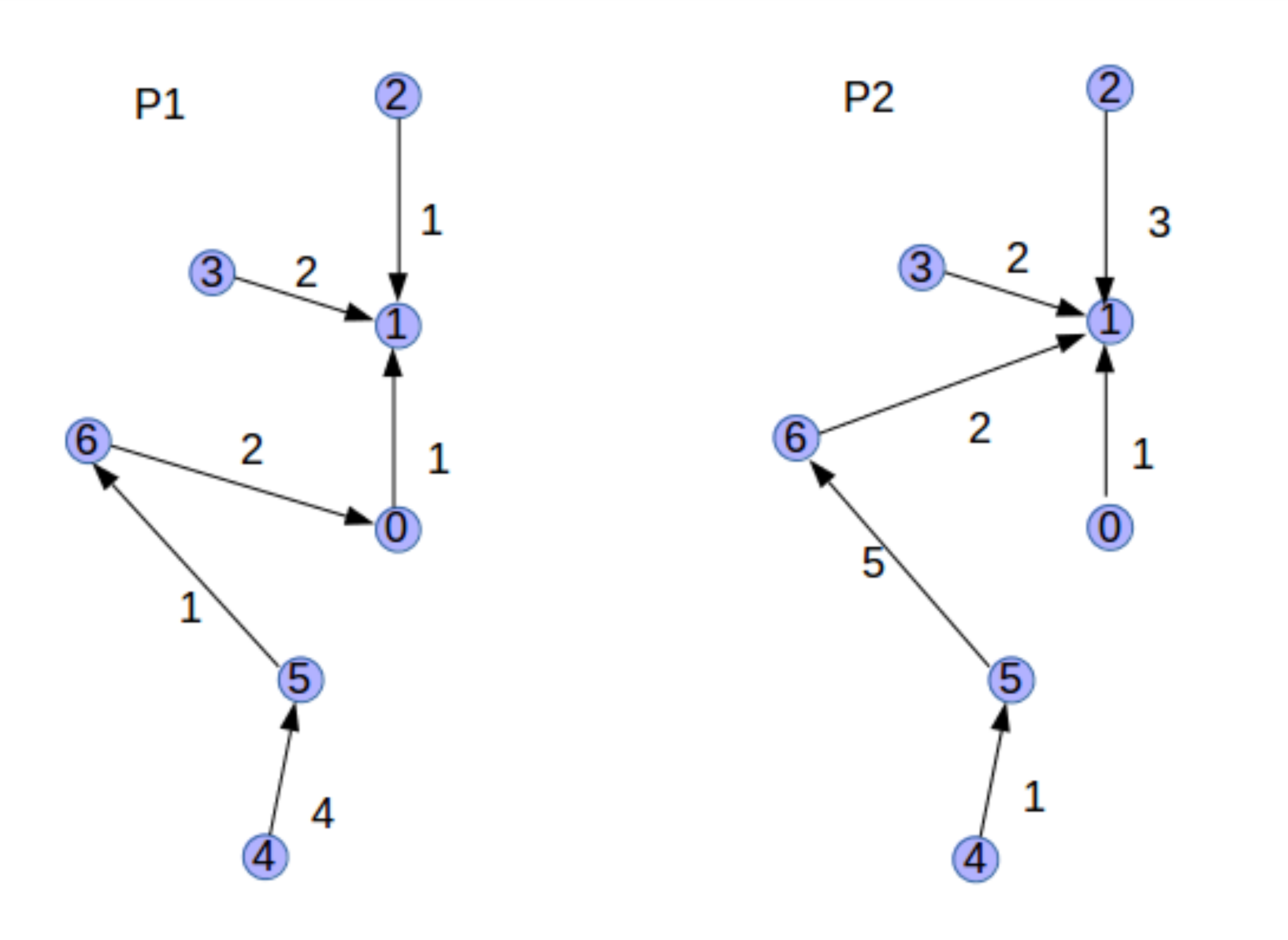, width=2.5in}
%\caption{Multi-Paths based strategies}
%\label{ExecutionTime}

\centering
\epsfig{file=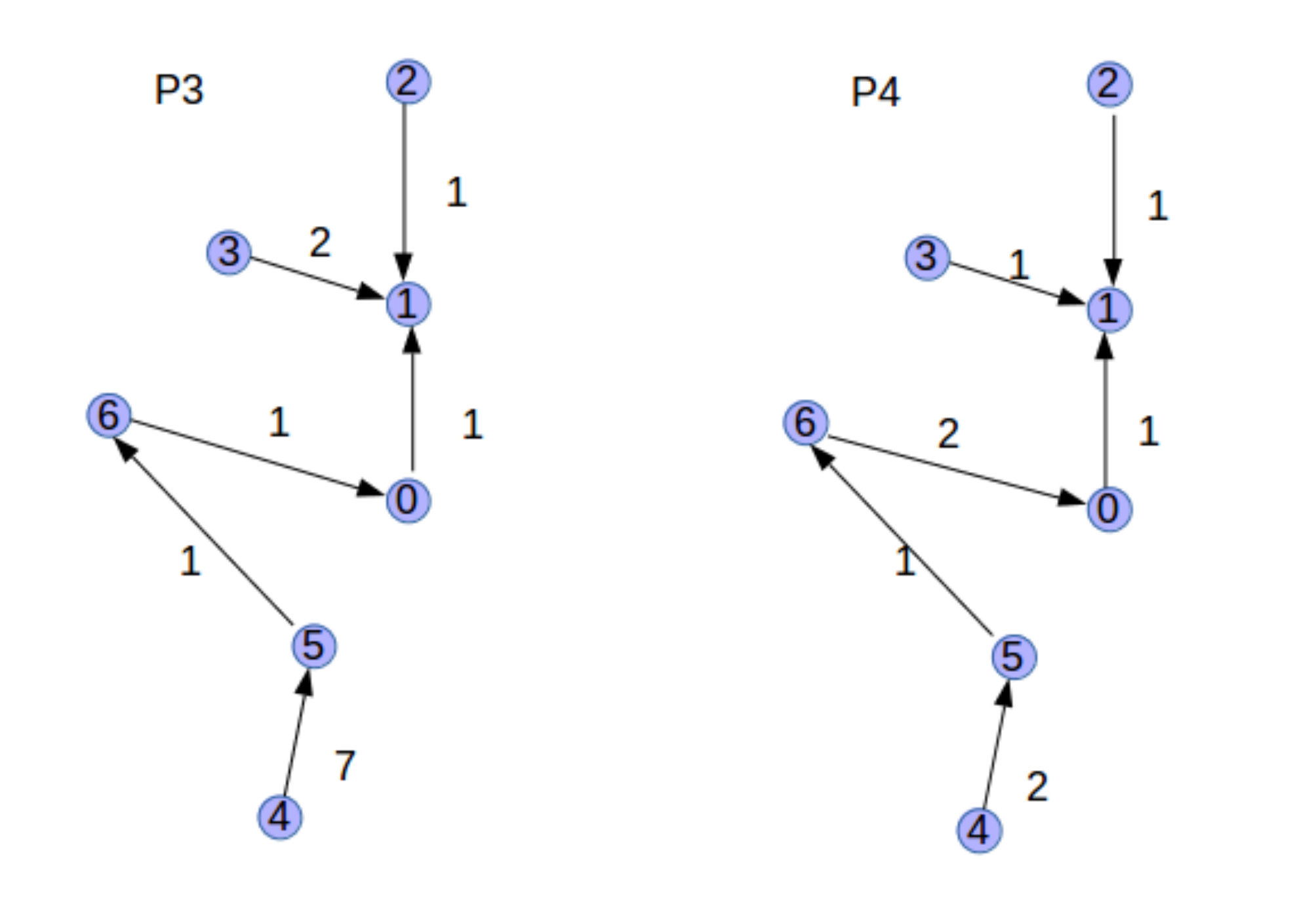, width=2.5in}
%\caption{Multi-Paths based strategies}
%\label{ExecutionTime}

\centering
\epsfig{file=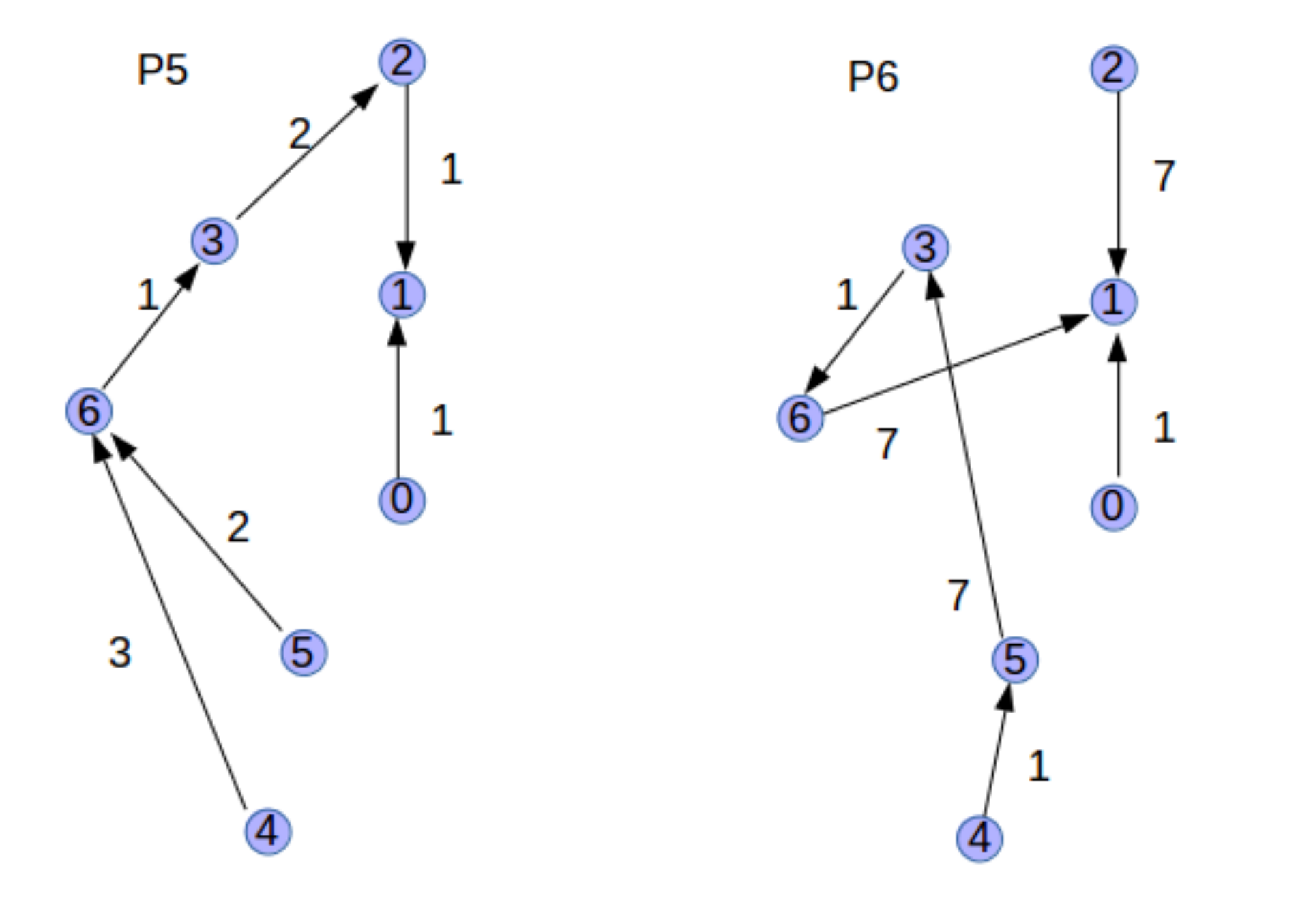, width=2.5in}
%\caption{Multi-Paths based strategies}
%\label{ExecutionTime}

\centering
\epsfig{file=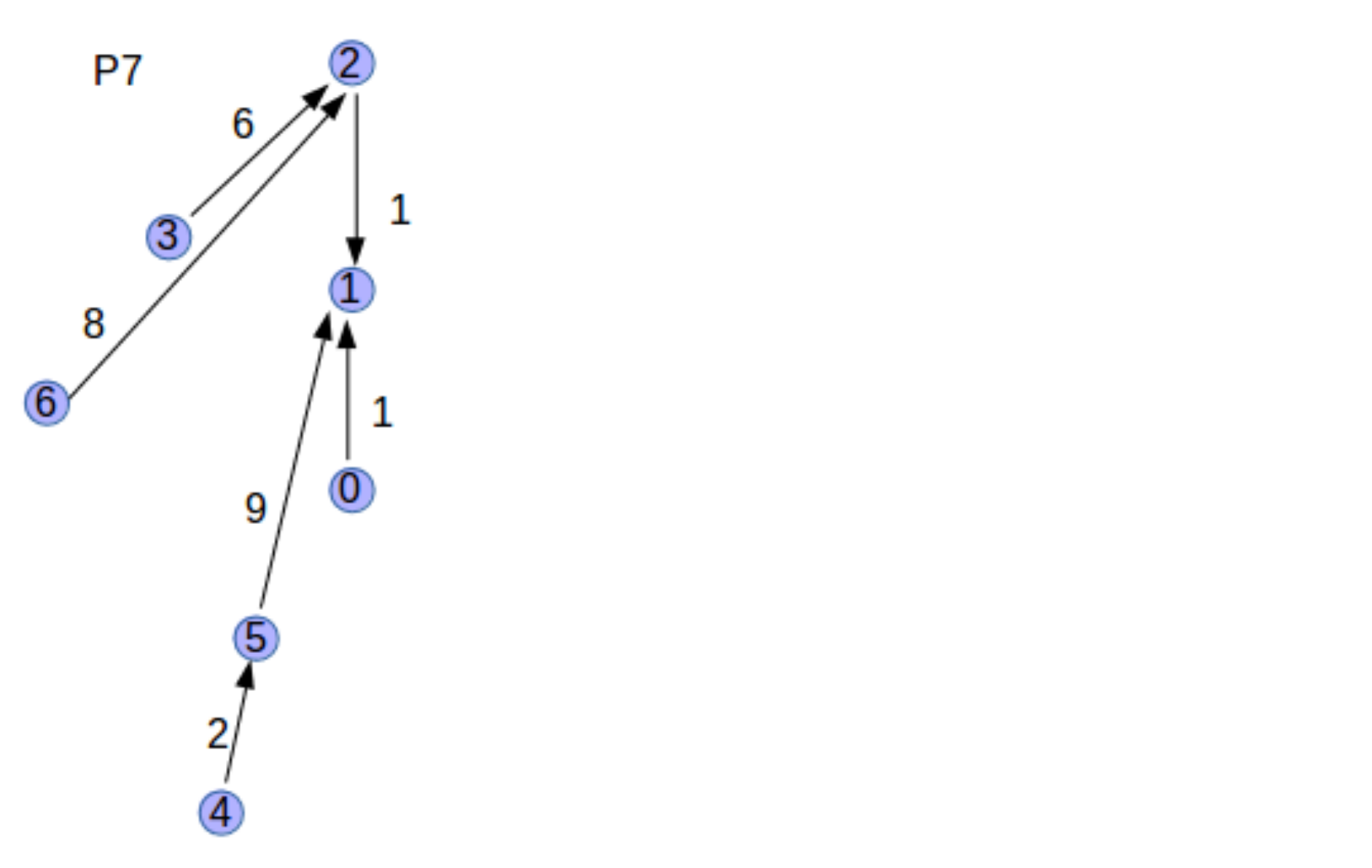, width=2.5in}
\caption{Different PPVG Trees in different Postures}
%\label{ExecutionTime}
\label{Fig10}
\end{figure}

In this section we propose a new convergecast tree based strategy, Probabilist Posture Varying Graph (PPVG). 
%Given this new WBAN model and the theory of No-ACK Retransmission, in this section, we give a new Milti-Hop convergecast Networks strategy witch is well adapted to the WBAN environment.
%
%\subsection{Best Path}
%
%Based on the defect of current network strategies, the new one should meet the three following characteristics: 
%
%1, Try to reduce the useless traffic transmission and number of hops. For example transmission of ACK and extra retransmission caused by the missing ACK.
%
%2, Try to keep one single path for each source to send packets. According to previous conclusions, the single path transmission could avoid the wrong-sequence of packets and if we can use No-ACK Retransmission to improve reliability, we don't need to use the Multi-Paths to do that.
%
%3, Try to keep a balance traffic for each relay node. Because of the Multi-Hop transmission, we hope to relatively spread traffic into all the network, not to make it all block in one single node, if possible.
%
%Following these three rulers, 
%We present a PPVG based Tree topology as a convergecast strategy to send and forward packets to the sink from sources. 
The idea is that each source has only one preselected parent to send traffic to and the parent forwards the traffic to its parent until it reaches the sink. For each transmission, we use No-ACK retransmission to ensure the reliability. The preselected path is calculated from connexion existing probability (see Figure \ref{Fig6}) and ETX of each transmission for a link (inverse of the probability). A preselected path is tree-based multi-hops path (sink as the root) for each source node to sink. The preselected path shows also the smallest total $ETX$ along a rout to sink for each source node. We have thus 7 different PPVG Trees for each posture (see Figure \ref{Fig10}). The number on each link is the $ETX$ for each link; the arrow indicates the parent.

%\subsection{Posture Detecting}

%Because the sensors we used have the capacity of detecting in which posture we are in real time, so our PVG Tree strategy can load different Tree topology according to the posture detecting result to better adapter to the WBAN model.

%However, for those sensors who have no capacity of detecting the posture, we can also get a solution from PPVG Tree: a General Tree. The General Tree is from the means probability of different postures for each link and it can have one general view for the posture. Further, we can decide with which percentage every posture link probability should be considered in the computing of the mean valuer.

%Obviously, using the General Tree is less better than using different PPVG Tree for corresponding posture, but it's also interesting for those function-simple sensors.

\begin{figure}[!t]
\centering
\epsfig{file=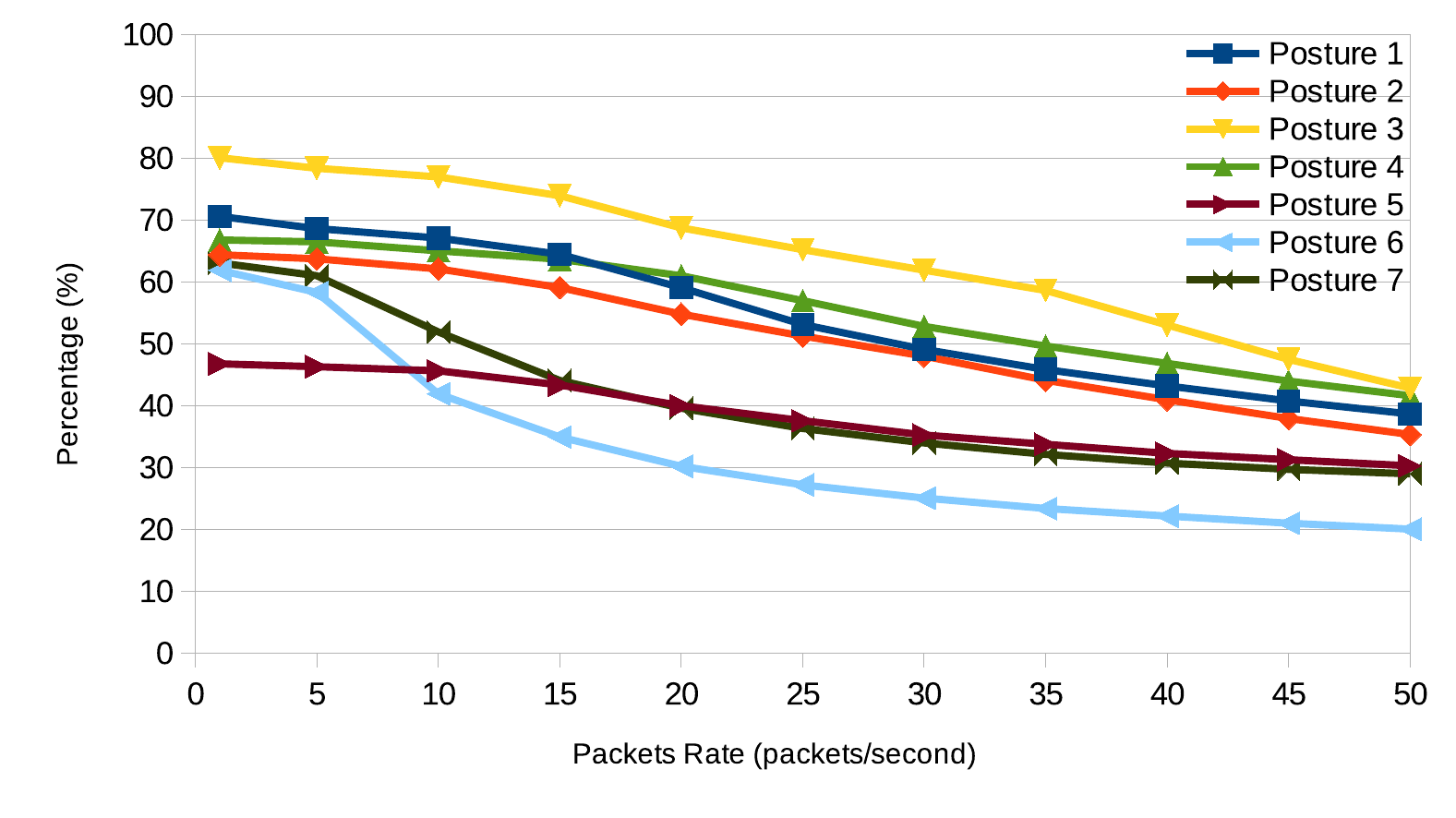, width=2.5in}
\caption{Reception Rate for PPVG Tree strategy in different Postures}
\label{Fig11}
\end{figure}

%\subsection{Simulation Result}
In the sequel we simulate the new PPVG Tree strategy using the same simulation environment as for the previous strategies. The simulation results are shown in Figure \ref{Fig11} for each posture, each of them takes the mean value over 10 simulation runs.

%\begin{figure}[!t]
%\centering
%\includegraphics[width=.9\columnwidth]{p17/p1a.png}
%\caption{PPVG Tree strategy in Posture 1}
%\label{Fig11}
%\end{figure}

%\begin{figure}[!t]
%\centering
%\includegraphics[width=.9\columnwidth]{p17/p2a.png}
%\caption{PPVG Tree strategy in Posture 2}
%\label{Fig12}
%\end{figure}

%\begin{figure}[!t]
%\centering
%\includegraphics[width=.9\columnwidth]{p17/p3a.png}
%\caption{PPVG Tree strategy in Posture 3}
%\label{Fig13}
%\end{figure}

%\begin{figure}[!t]
%\centering
%\includegraphics[width=.9\columnwidth]{p17/p4a.png}
%\caption{PPVG Tree strategy in Posture 4}
%\label{Fig14}
%\end{figure}

%\begin{figure}[!t]
%\centering
%\includegraphics[width=.9\columnwidth]{p17/p5a.png}
%\caption{PPVG Tree strategy in Posture 5}
%\label{Fig15}
%\end{figure}

%\begin{figure}[!t]
%\centering
%\includegraphics[width=.9\columnwidth]{p17/p6a.png}
%\caption{PPVG Tree strategy in Posture 6}
%\label{Fig16}
%\end{figure}

%\begin{figure}[!t]
%\centering
%\includegraphics[width=.9\columnwidth]{p17/p7a.png}
%\caption{PPVG Tree strategy in Posture 7}
%\label{Fig17}
%\end{figure}

The reception rate of our strategy is stable and slow-decreases except for posture 6 and 7, which present a clear decrease for  packets rate greater than 5 packets par second.

Also, our  simulation results show that our strategy preserves the total order (no wrong-sequence) for each posture.

\begin{figure}[!t]
\centering
\epsfig{file=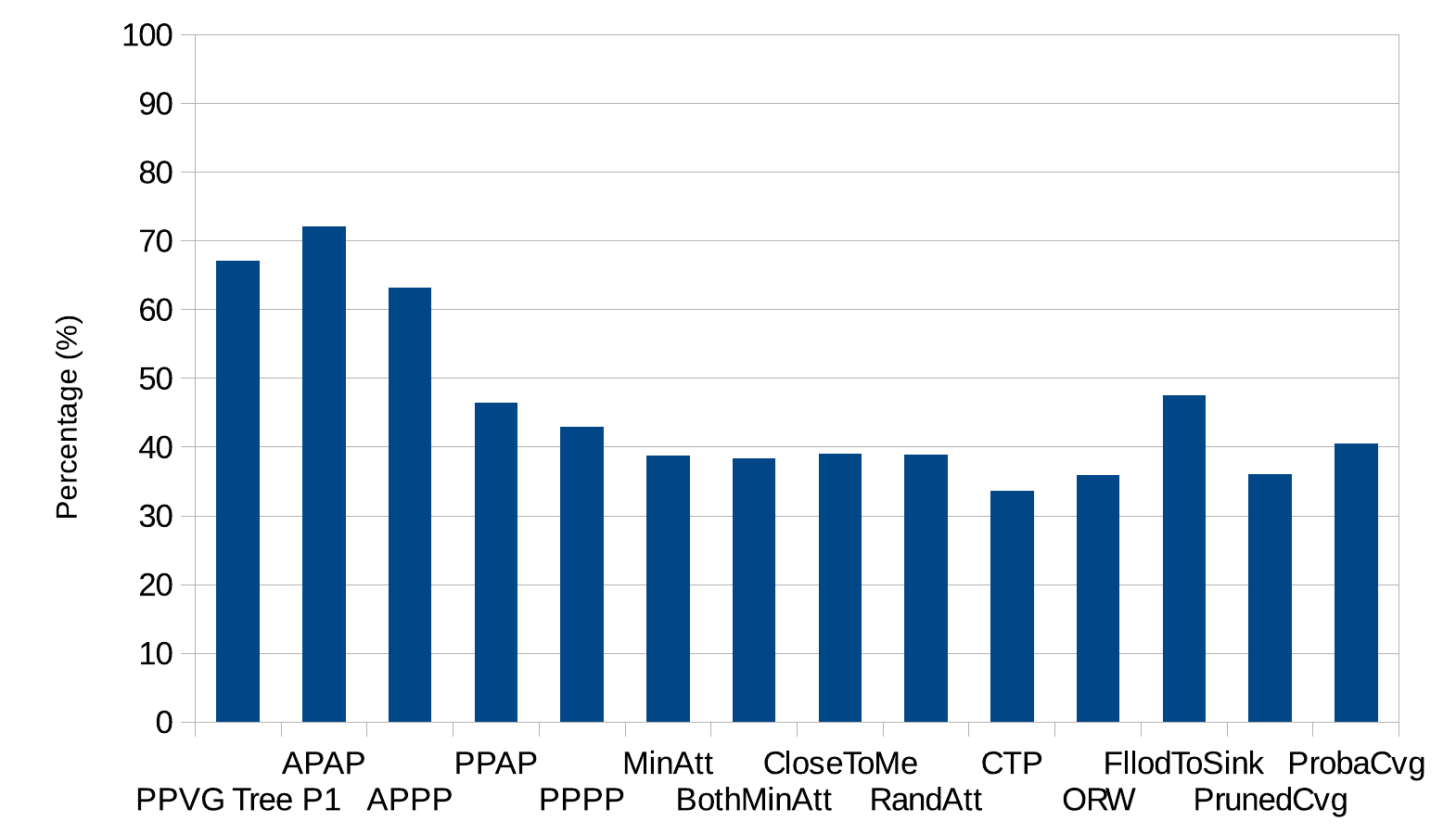, width=2.5in}
%\caption{Reception Rate in Posture 1 - 4}
%\label{Fig21}
%\end{figure}

%\begin{figure}[!t]
\centering
\epsfig{file=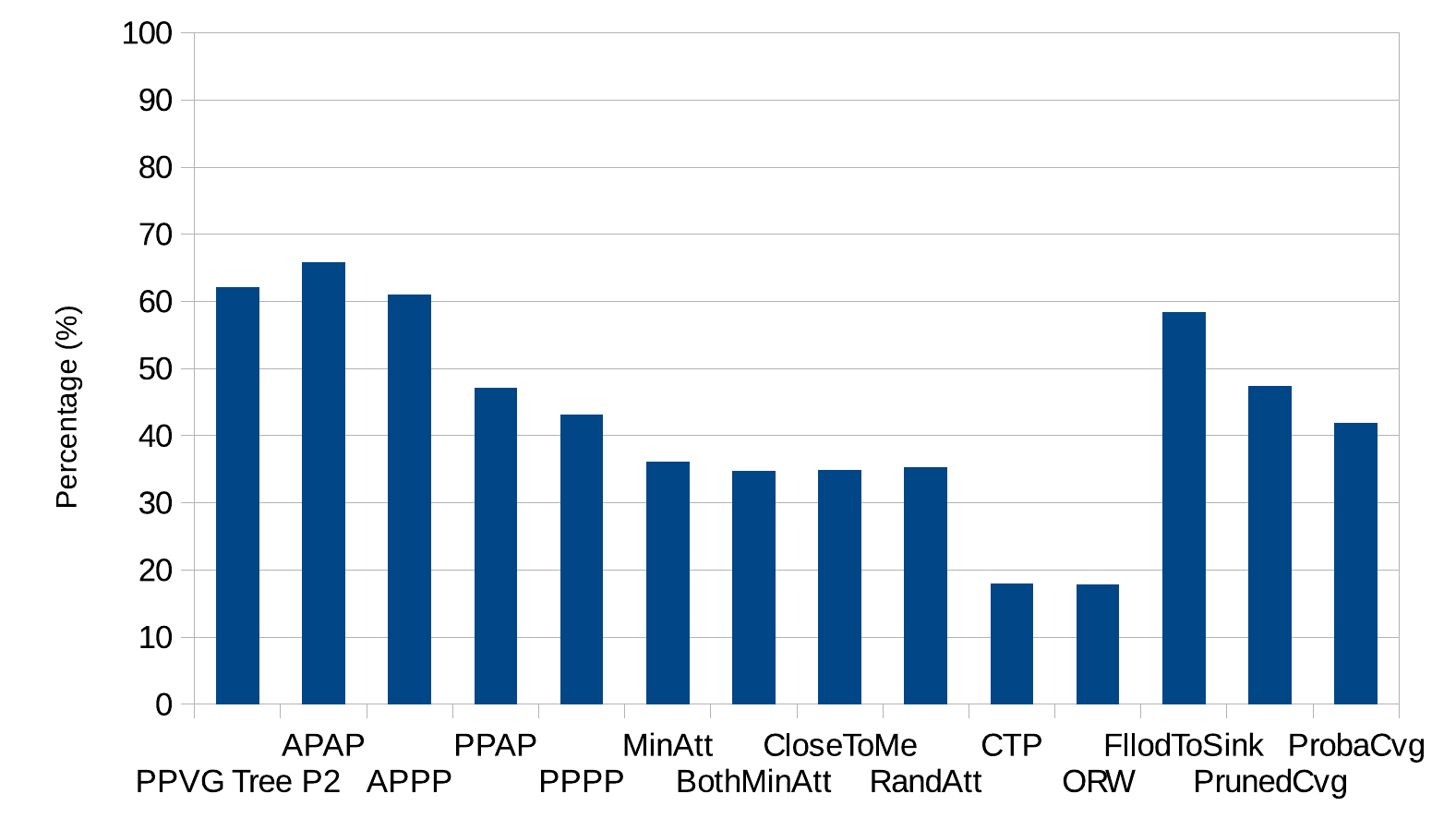, width=2.5in}
%\caption{Reception Rate in Posture 1 - 4}
%\label{Fig22}
%\end{figure}

%\begin{figure}[!t]
\centering
\epsfig{file=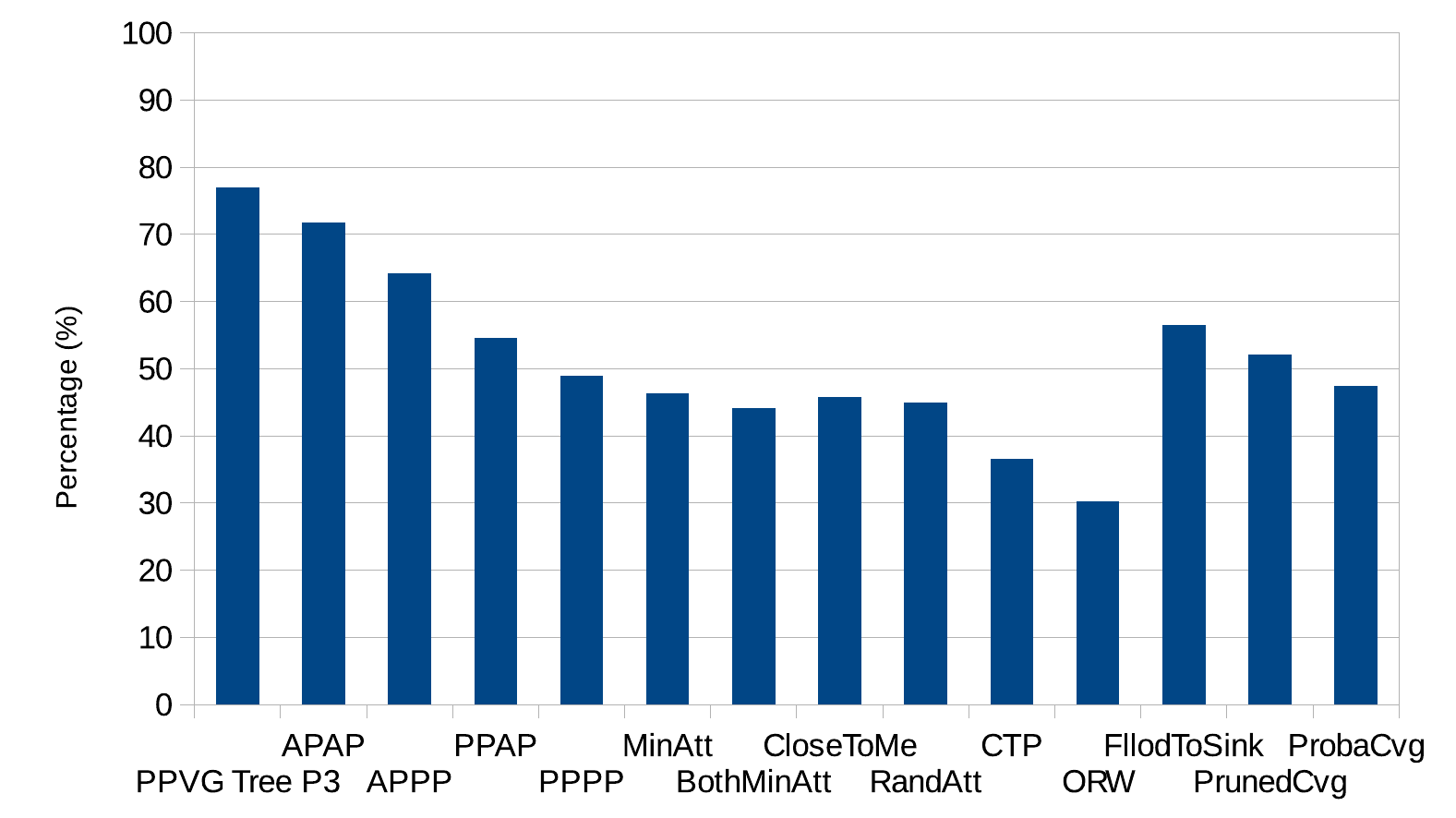, width=2.5in}
%\caption{Reception Rate in Posture 1 - 4}
%\label{Fig23}
%\end{figure}

%\begin{figure}[!t]
\centering
\epsfig{file=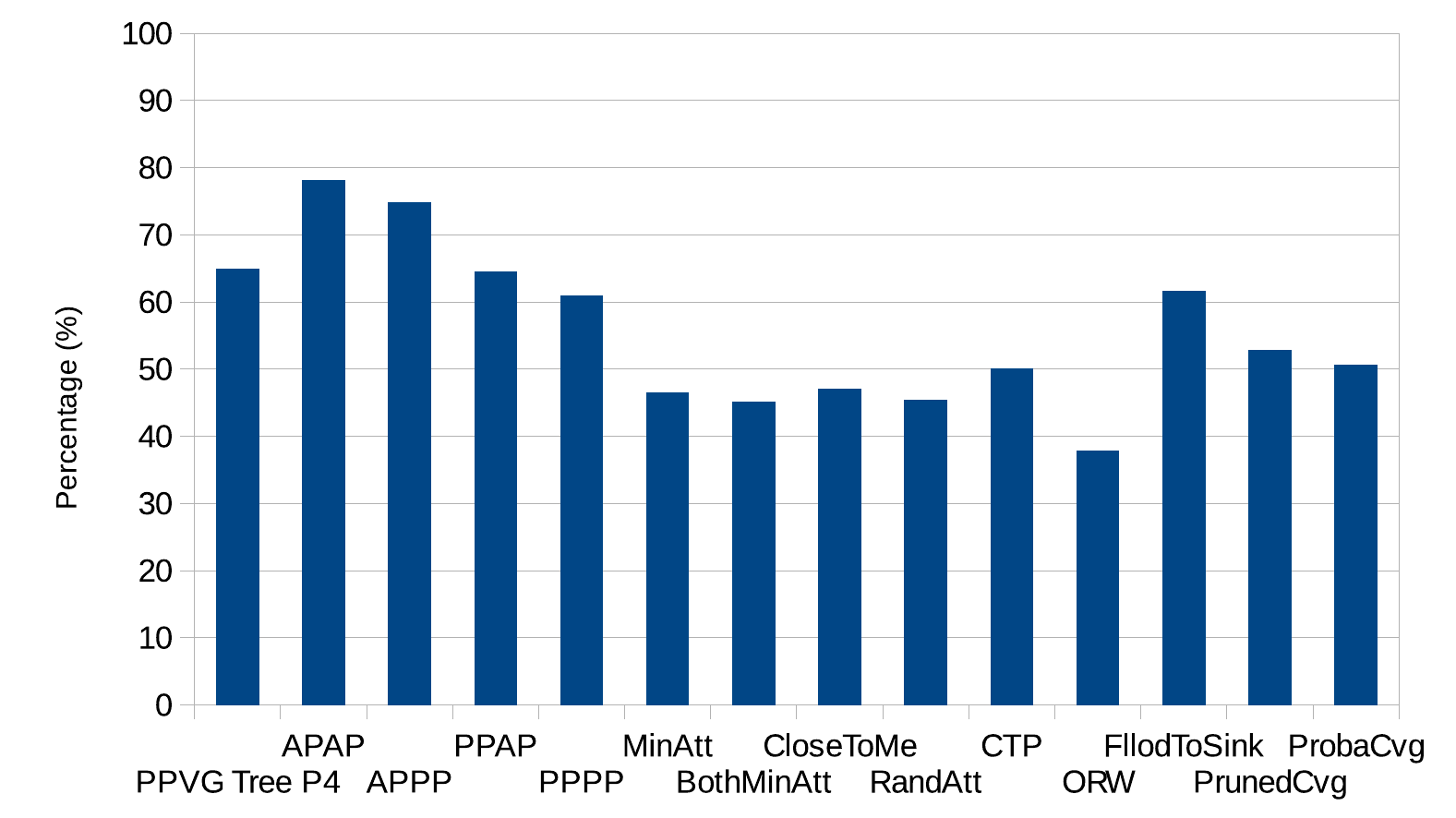, width=2.5in}
\caption{Reception Rate comparison among all the strategies in Posture 1-4}
\label{Fig24}
\end{figure}

\begin{figure}[!t]
\centering
\epsfig{file=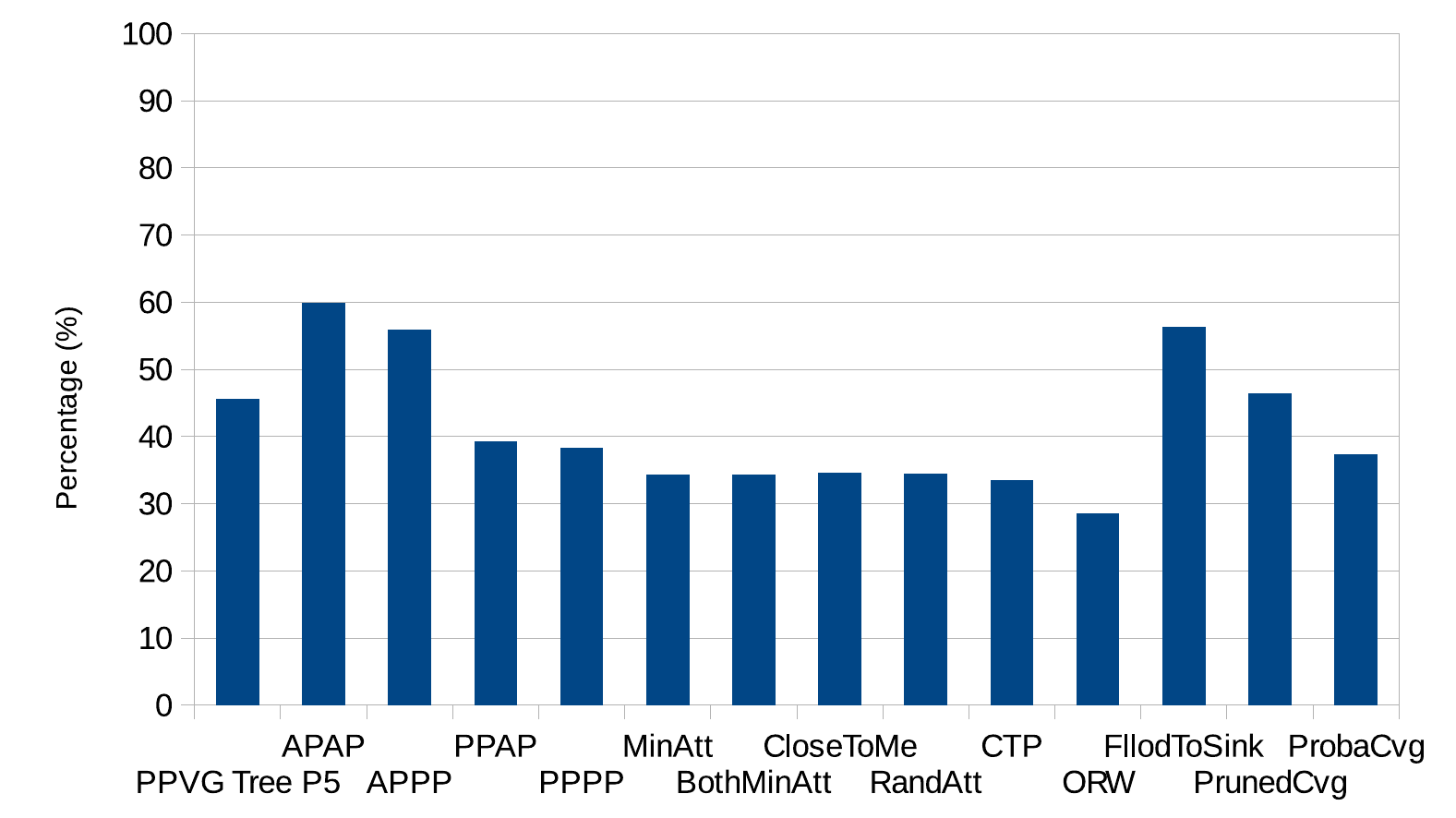, width=2.5in}
%\caption{Reception Rate in Posture 1 - 4}
%\label{Fig25}
%\end{figure}

%\begin{figure}[!t]
\centering
\epsfig{file=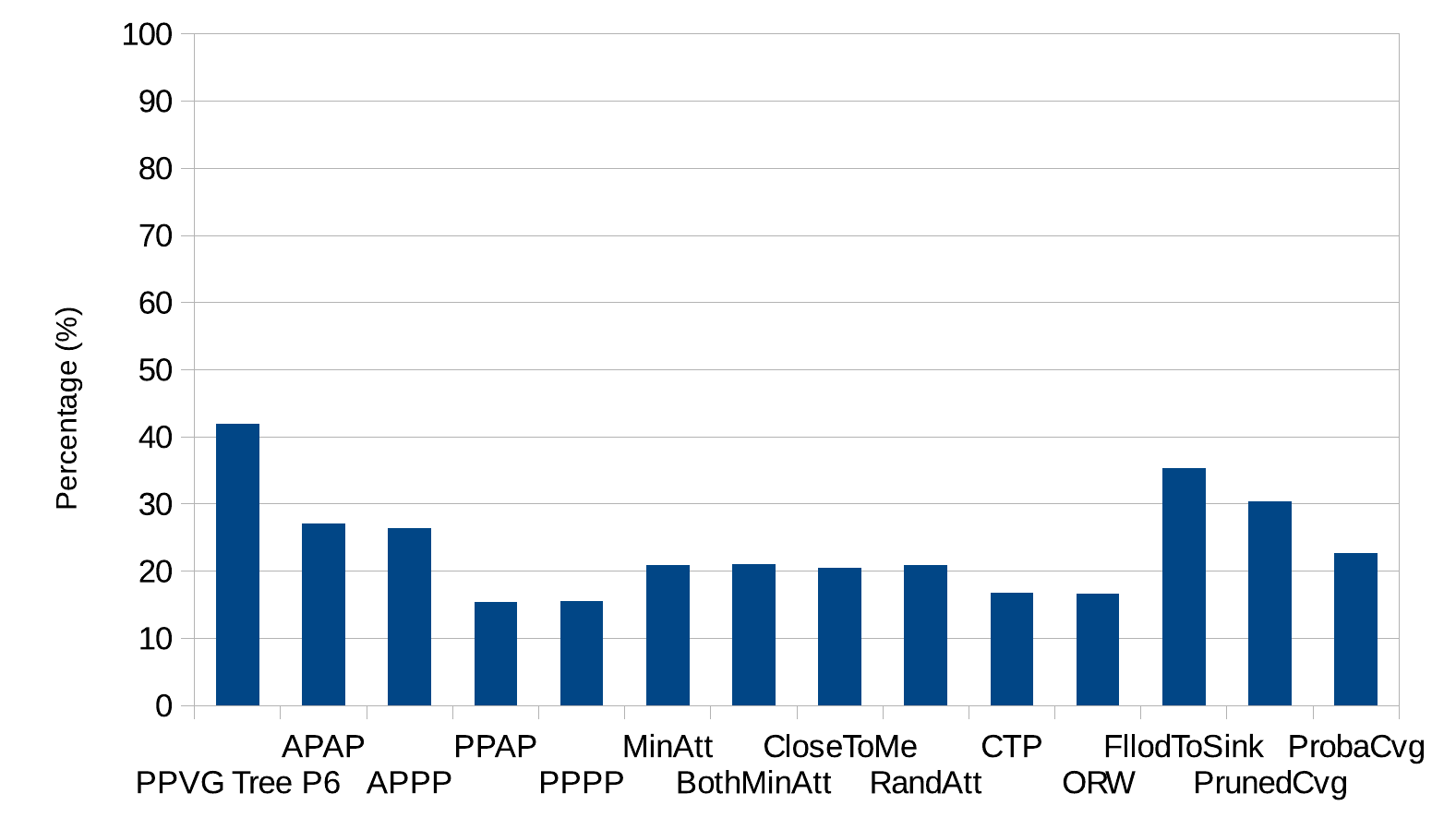, width=2.5in}
%\caption{Reception Rate in Posture 1 - 4}
%\label{Fig26}
%\end{figure}

%\begin{figure}[!t]
\centering
\epsfig{file=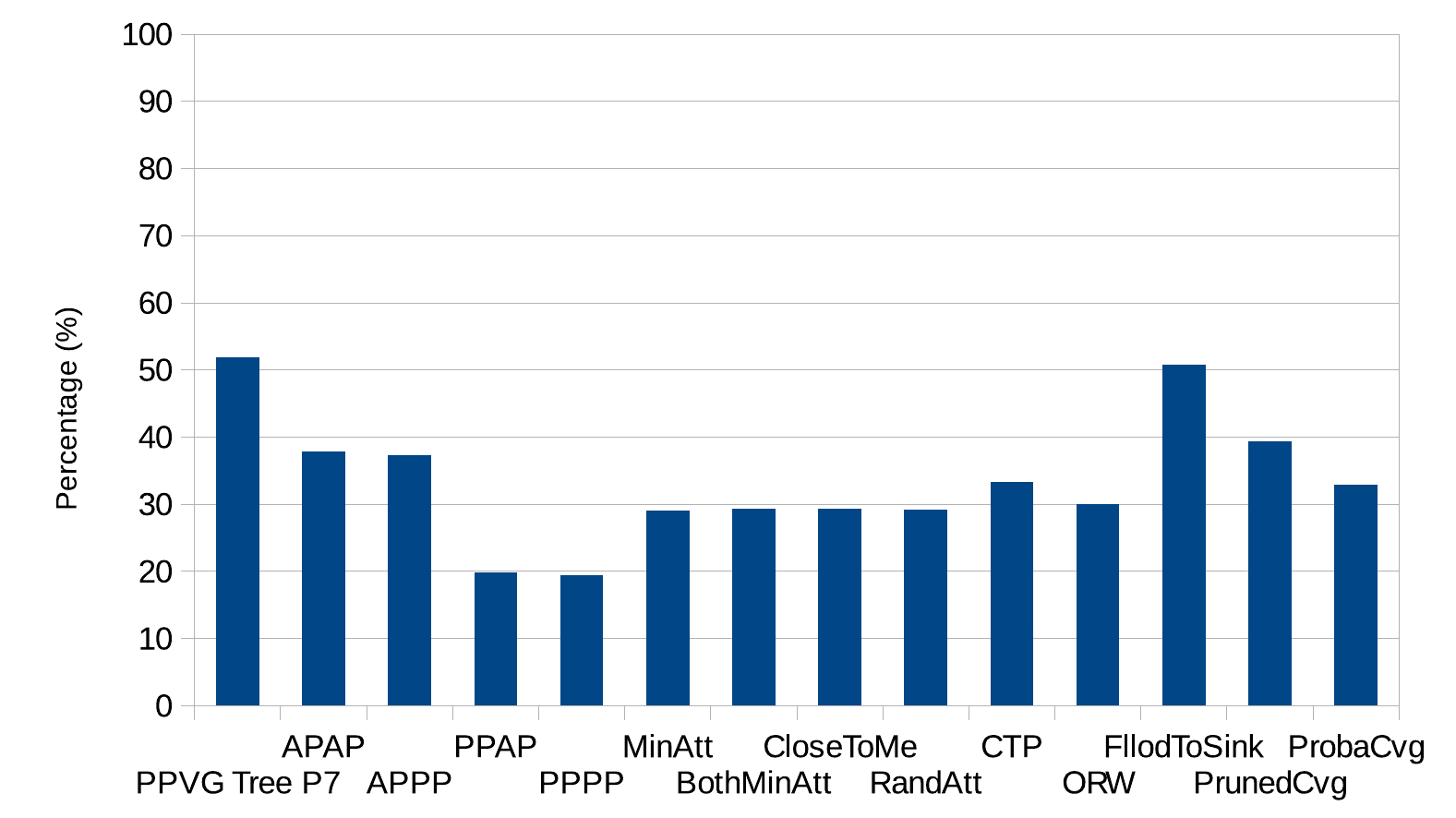, width=2.5in}
\caption{Reception Rate comparison among all the strategies in Posture 5-7}
\label{Fig27}
\end{figure}

\begin{figure}[!t]
\centering
\epsfig{file=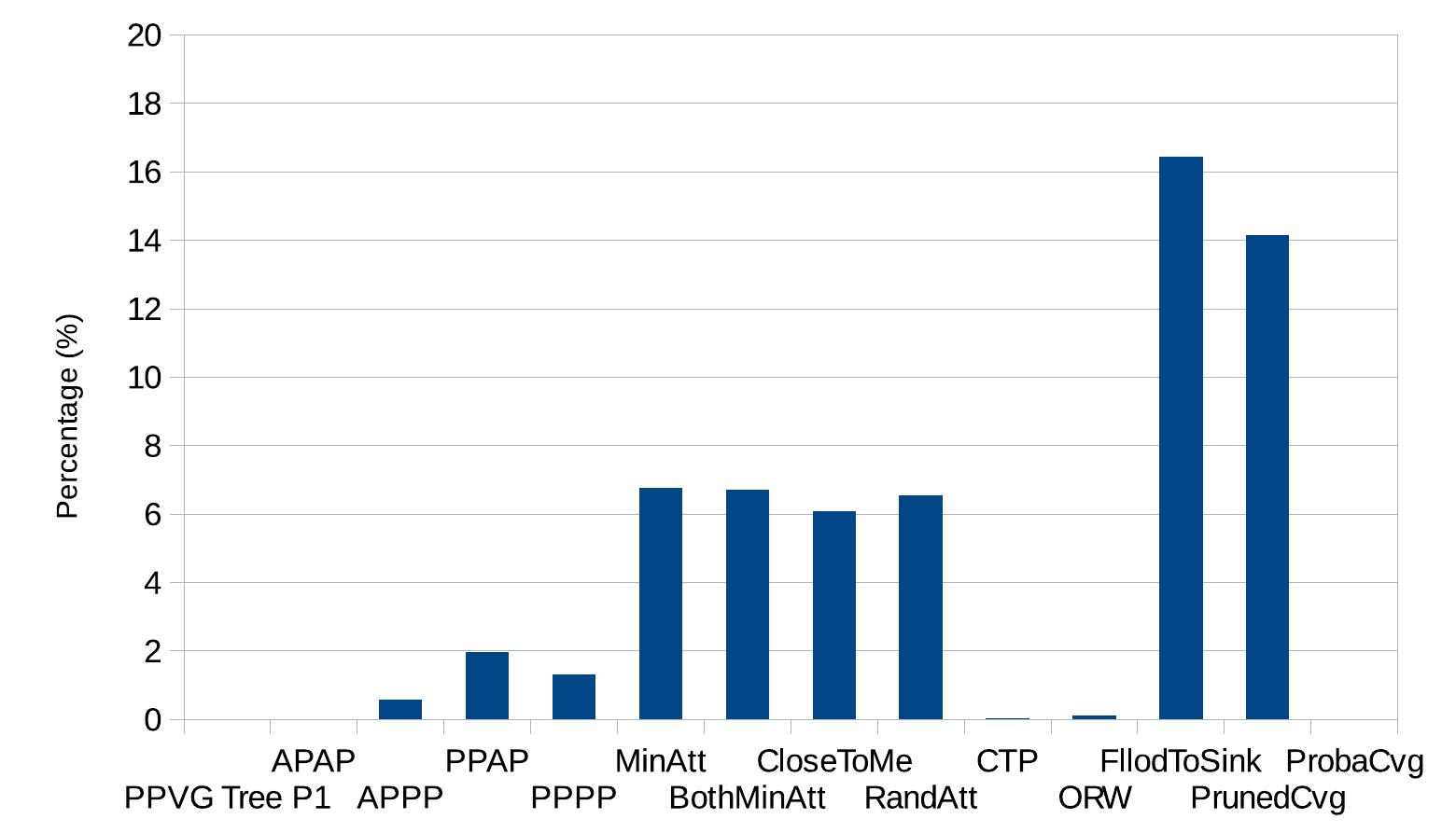, width=2.5in}
%\caption{Wrong-Sequence Rate in Posture 1 - 4}
%\label{Fig28}
%\end{figure}

%\begin{figure}[!t]
\centering
\epsfig{file=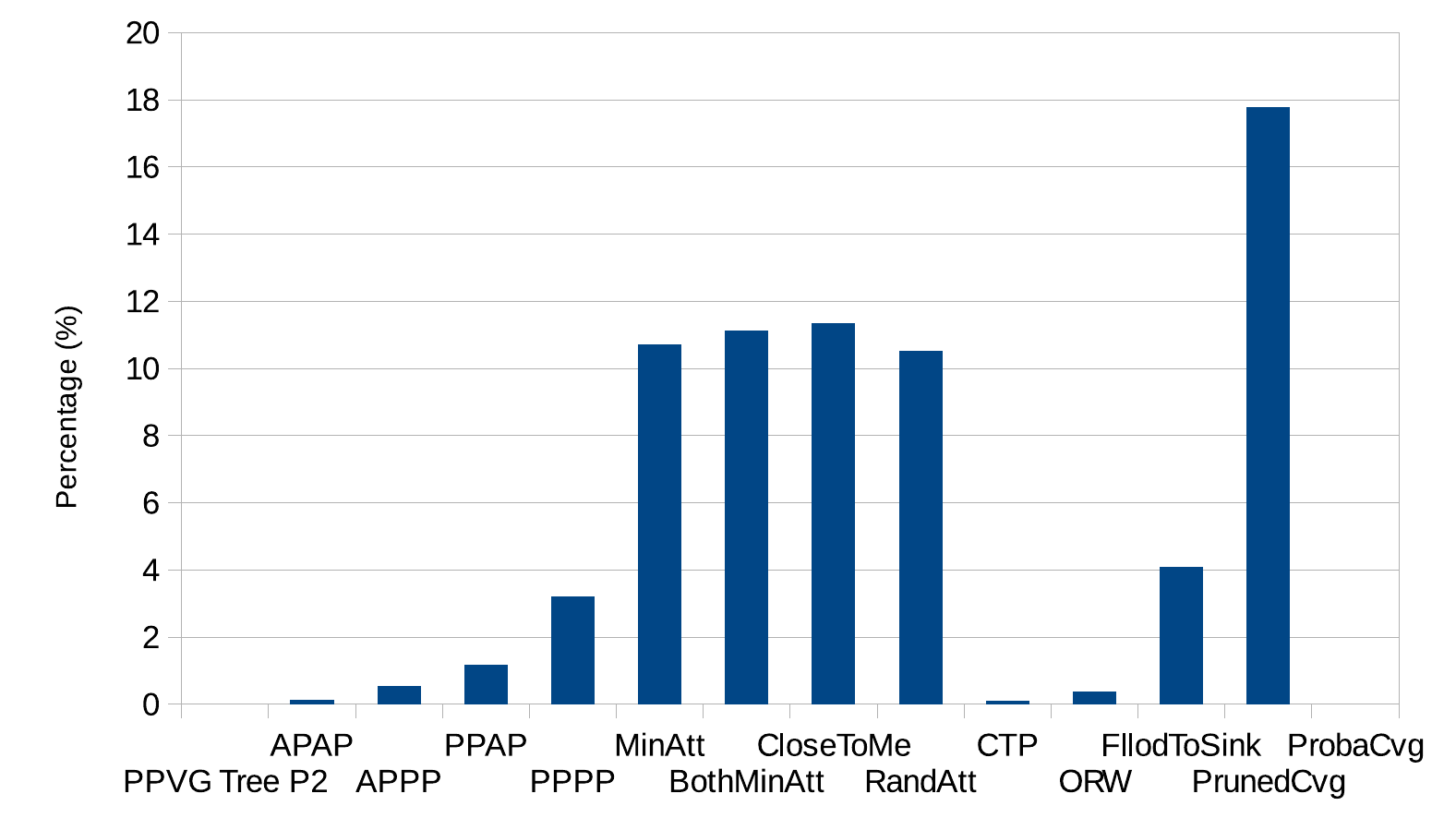, width=2.5in}
%\caption{Wrong-Sequence Rate in Posture 1 - 4}
%\label{Fig29}
%\end{figure}

%\begin{figure}[!t]
\centering
\epsfig{file=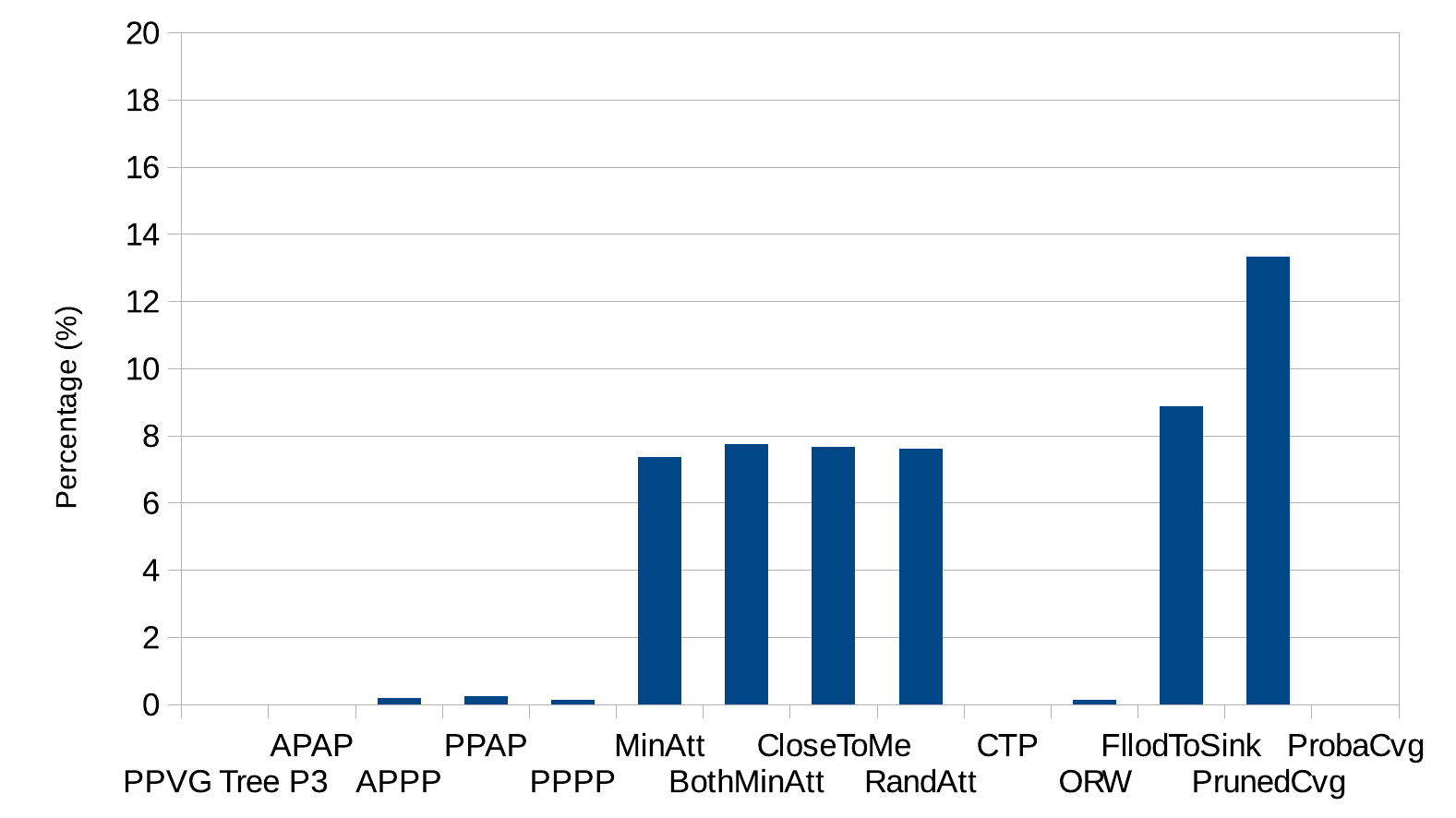, width=2.5in}
%\caption{Wrong-Sequence Rate in Posture 1 - 4}
%\label{Fig30}
%\end{figure}

%\begin{figure}[!t]
\centering
\epsfig{file=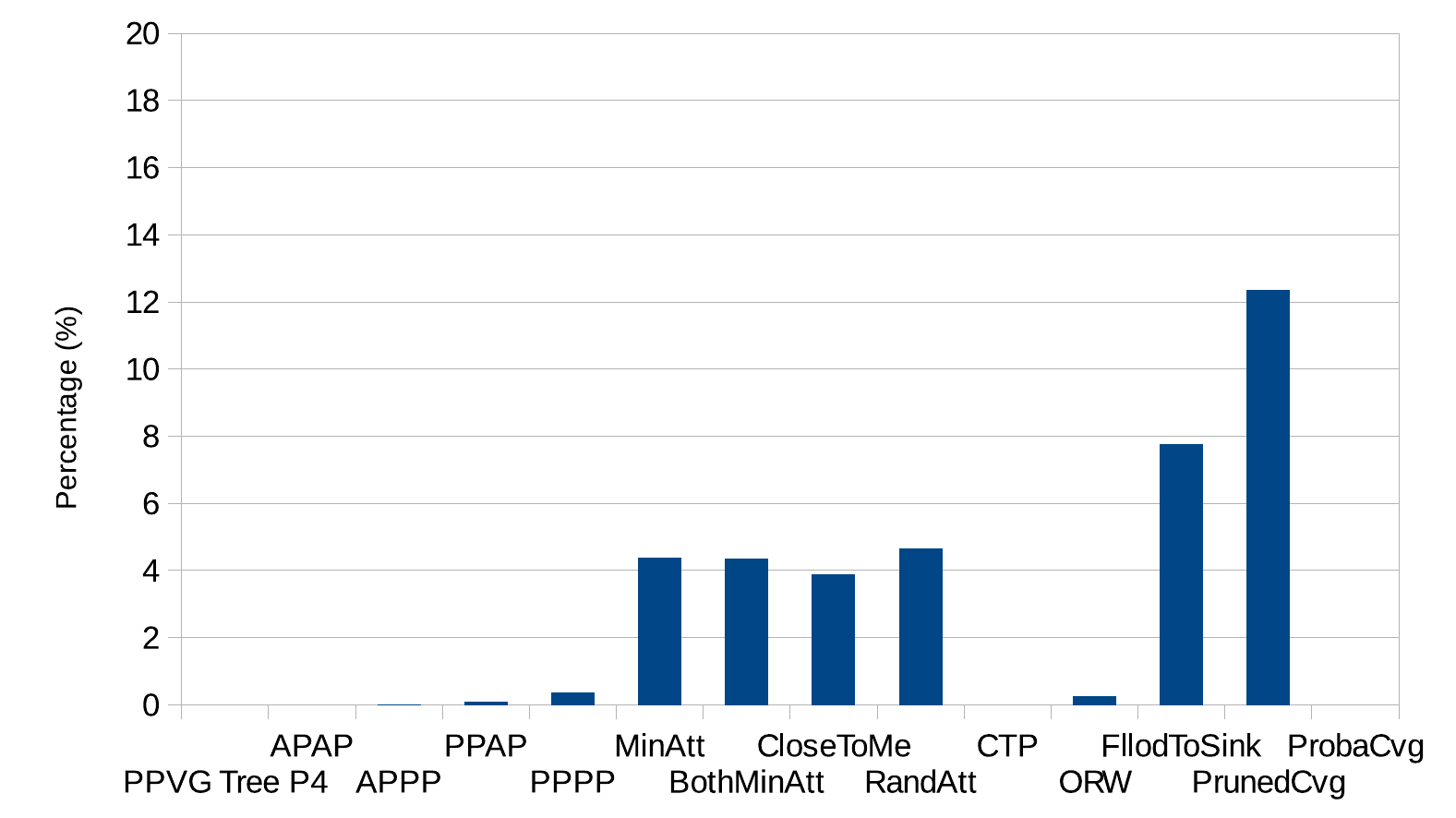, width=2.5in}
\caption{Packets Inversion Rate comparison among all the strategies in Posture 1-4}
\label{Fig31}
\end{figure}

\begin{figure}[!t]
\centering
\epsfig{file=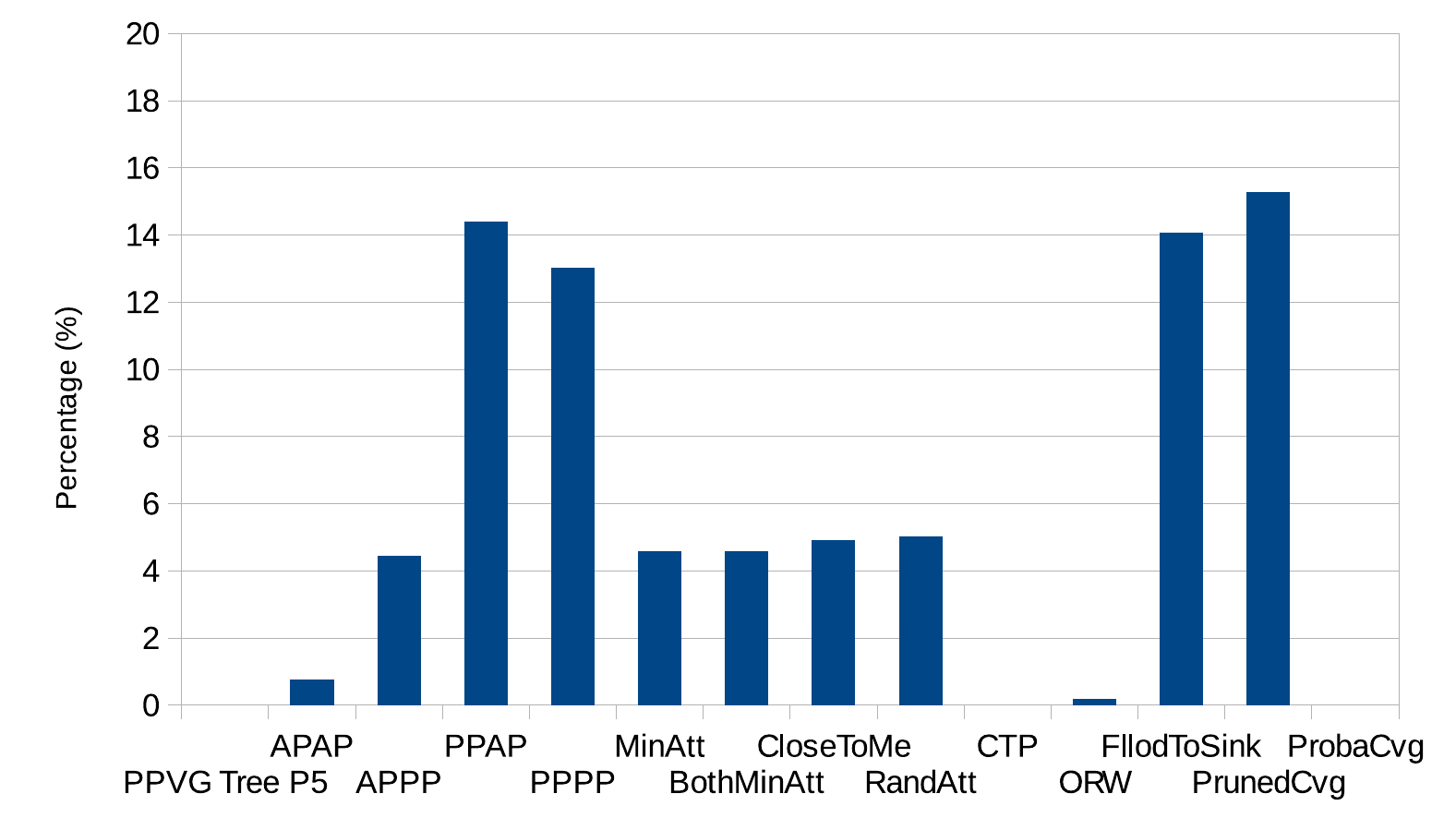, width=2.5in}
%\caption{Wrong-Sequence Rate in Posture 1 - 4}
%\label{Fig32}
%\end{figure}

%\begin{figure}[!t]
\centering
\epsfig{file=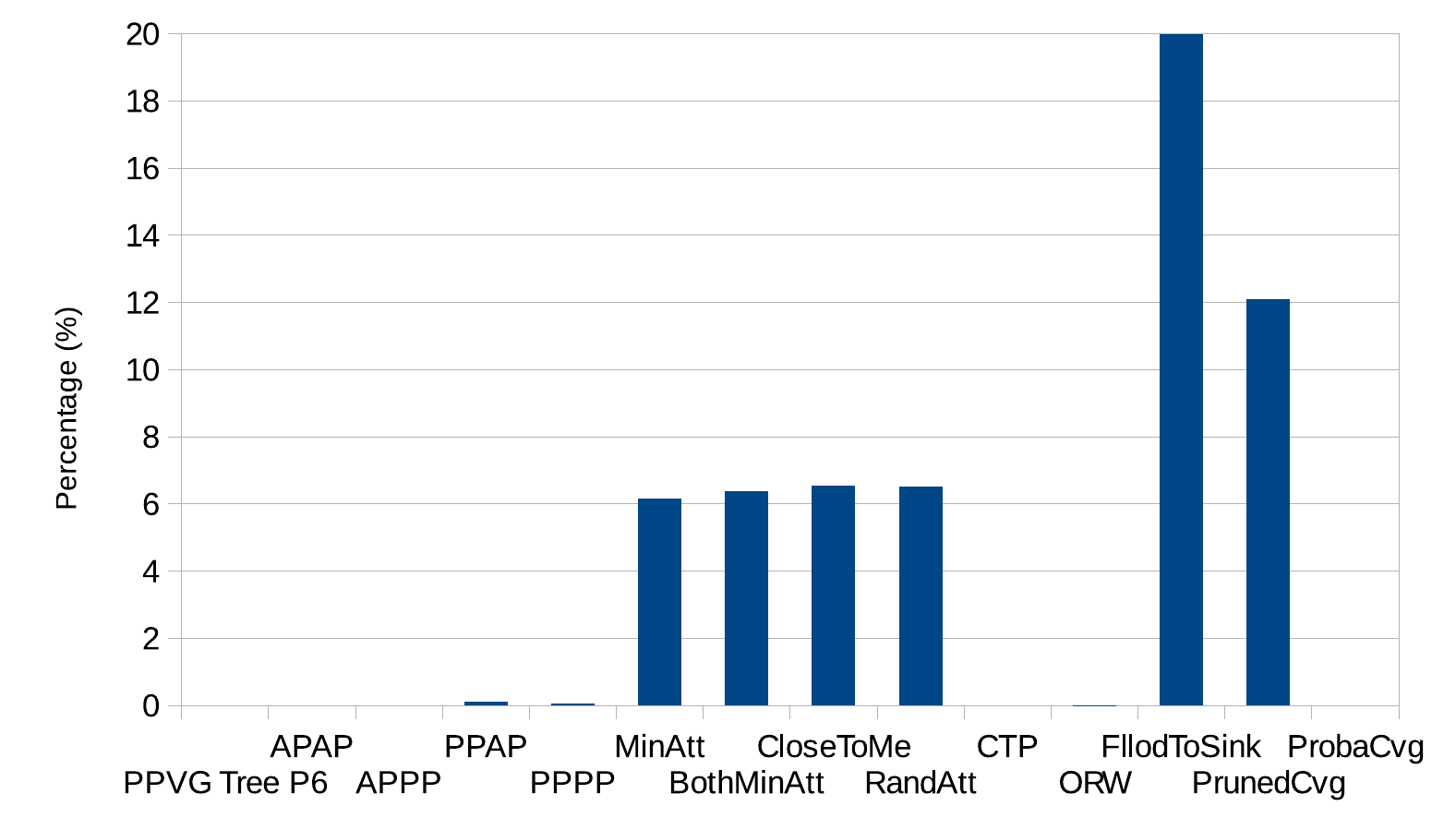, width=2.5in}
%\caption{Wrong-Sequence Rate in Posture 1 - 4}
%\label{Fig33}
%\end{figure}

%\begin{figure}[!t]
\centering
\epsfig{file=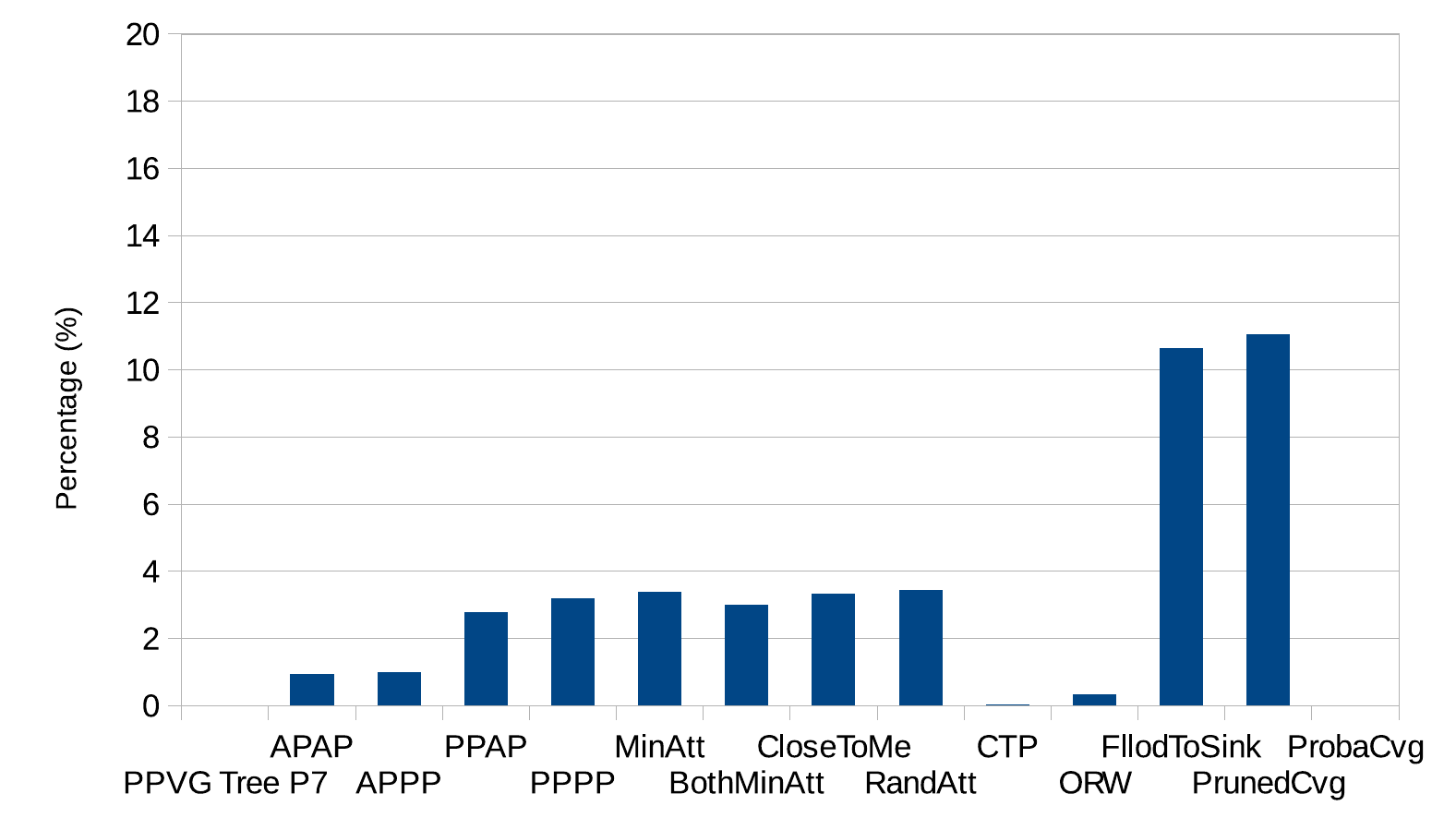, width=2.5in}
\caption{Packets Inversion Rate comparison among all the strategies in Posture 5-7}
\label{Fig34}
\end{figure}

\begin{figure}[!t]
\centering
\epsfig{file=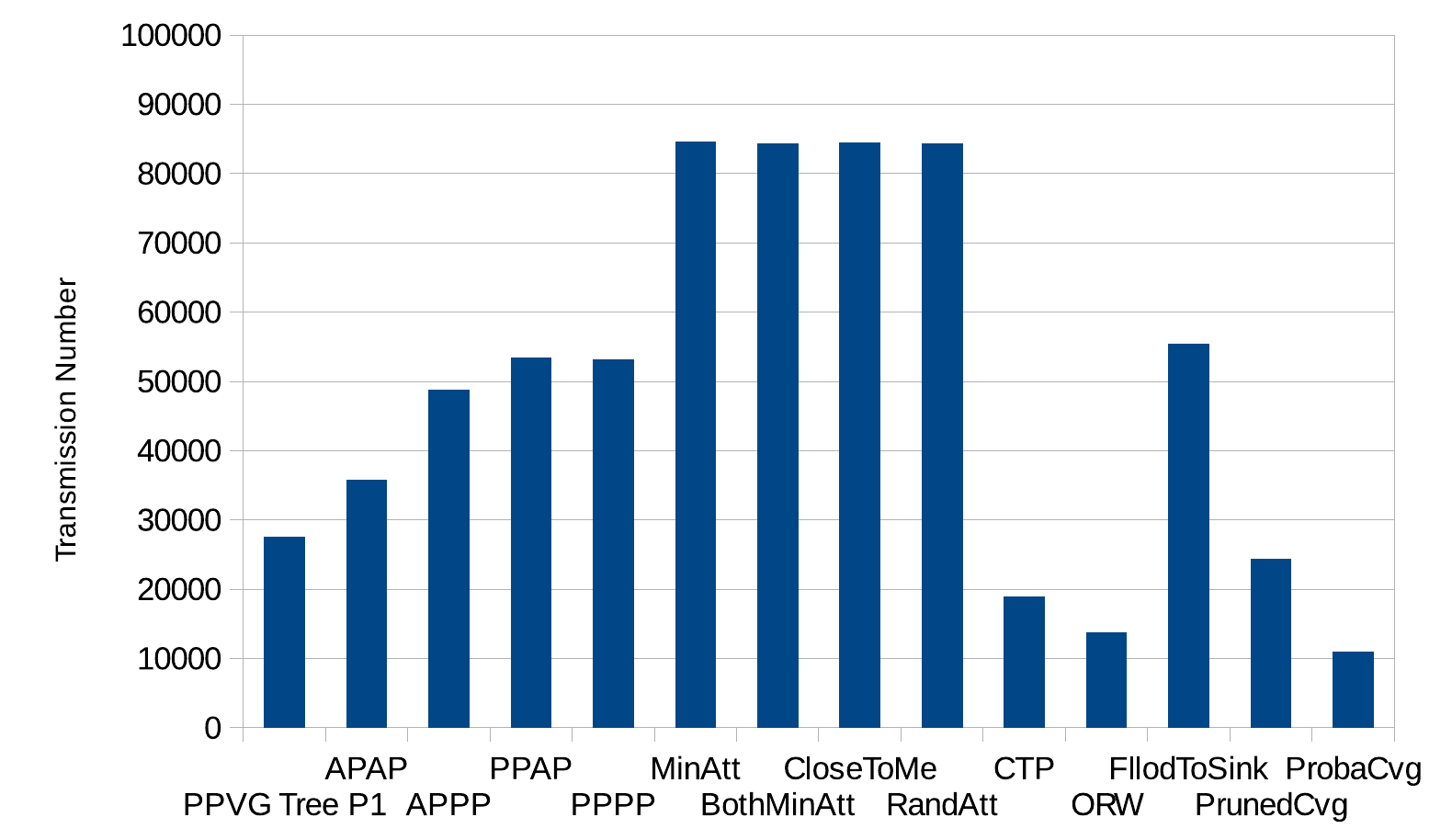, width=2.5in}
%\caption{Number of Transmissions in Posture 1- 4}
%\label{Fig35}
%\end{figure}

%\begin{figure}[!t]
\centering
\epsfig{file=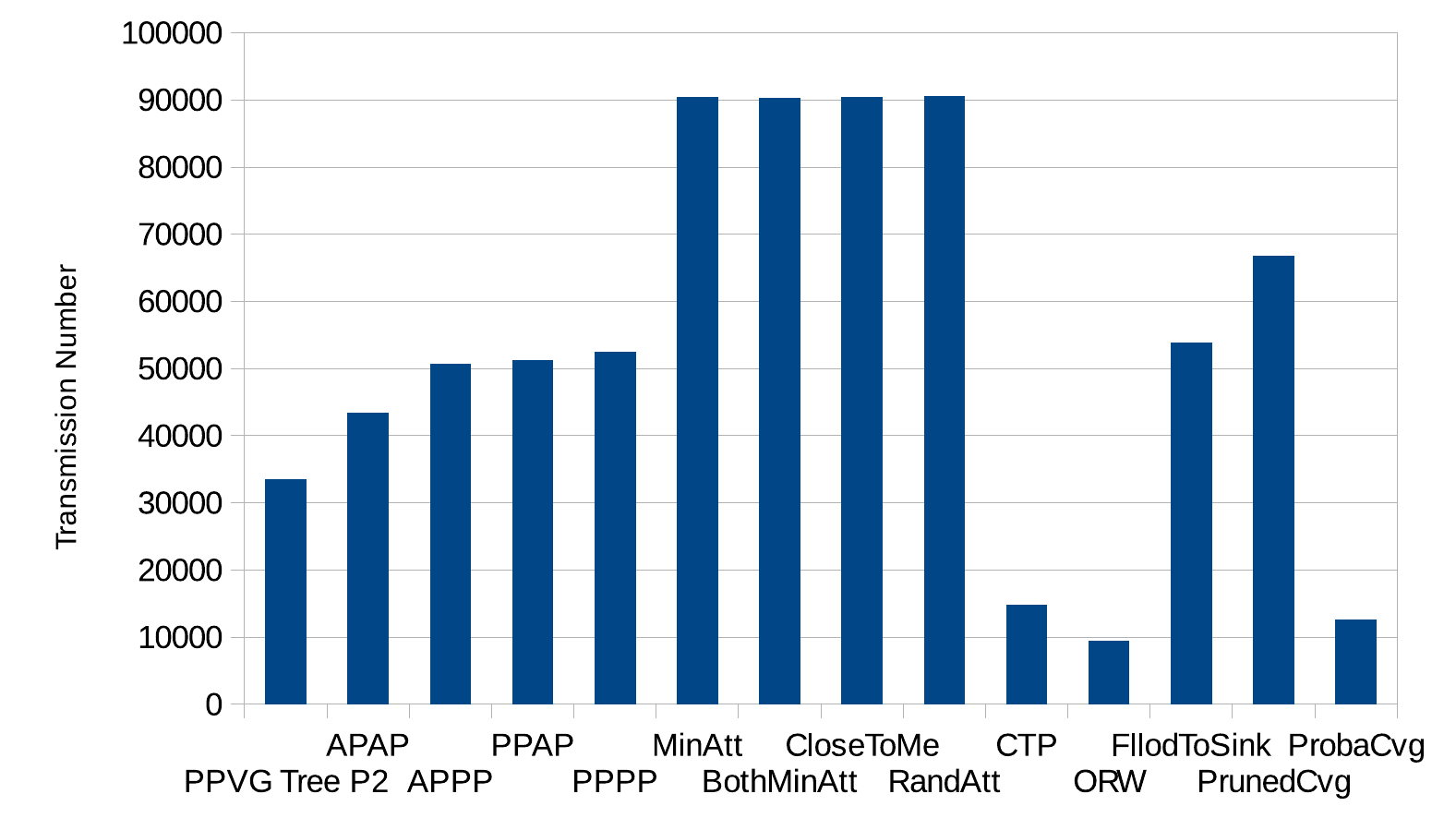, width=2.5in}
%\caption{Number of Transmissions in Posture 1- 4}
%\label{Fig36}
%\end{figure}

%\begin{figure}[!t]
\centering
\epsfig{file=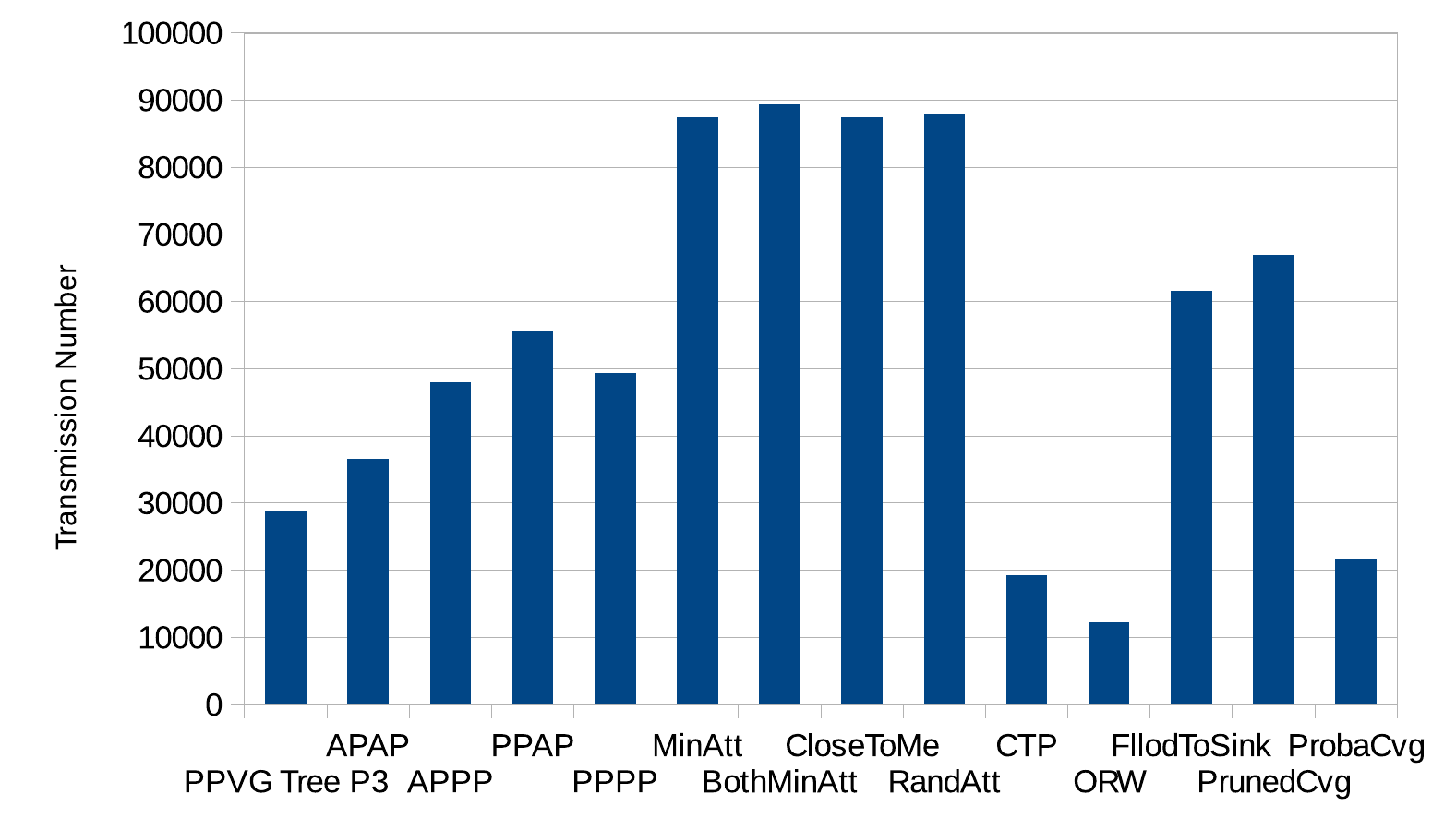, width=2.5in}
%\caption{Number of Transmissions in Posture 1- 4}
%\label{Fig37}
%\end{figure}

%\begin{figure}[!t]
\centering
\epsfig{file=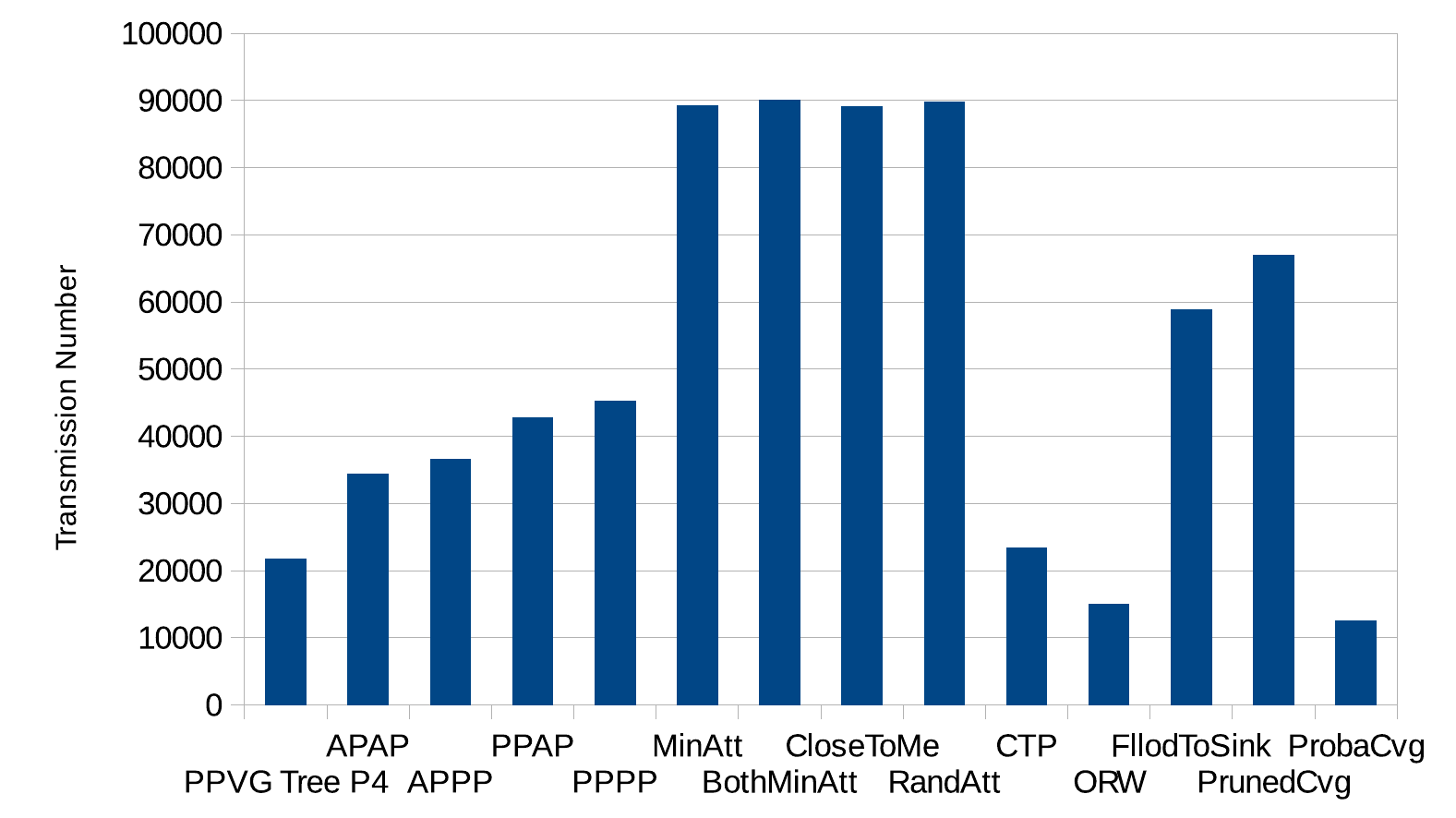, width=2.5in}
\caption{Number of Transmissions comparison among all the strategies in Posture 1-4}
\label{Fig38}
\end{figure}

\begin{figure}[!t]
\centering
\epsfig{file=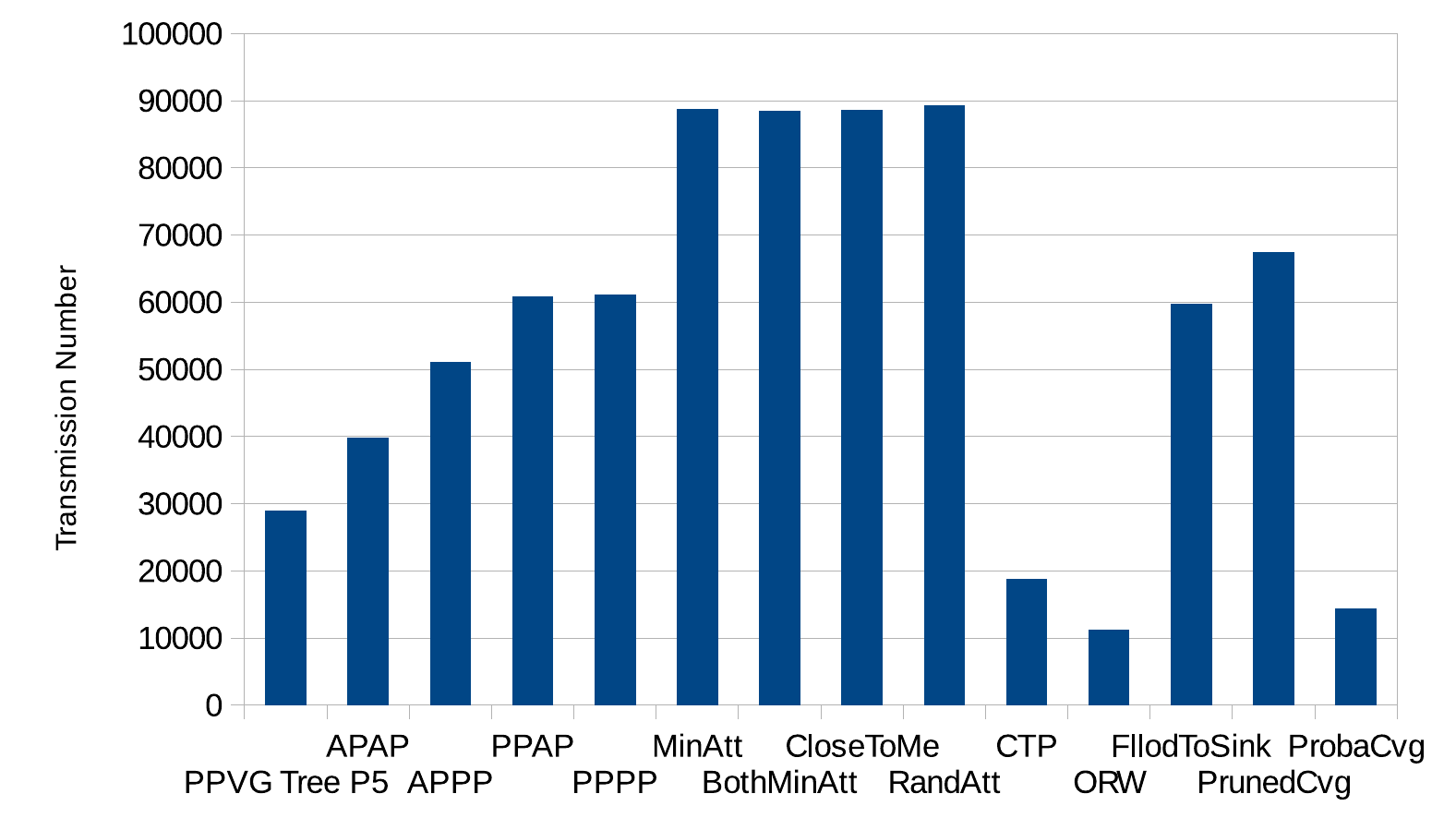, width=2.5in}
%\caption{Number of Transmissions in Posture 1- 4}
%\label{Fig39}
%\end{figure}

%\begin{figure}[!t]
\centering
\epsfig{file=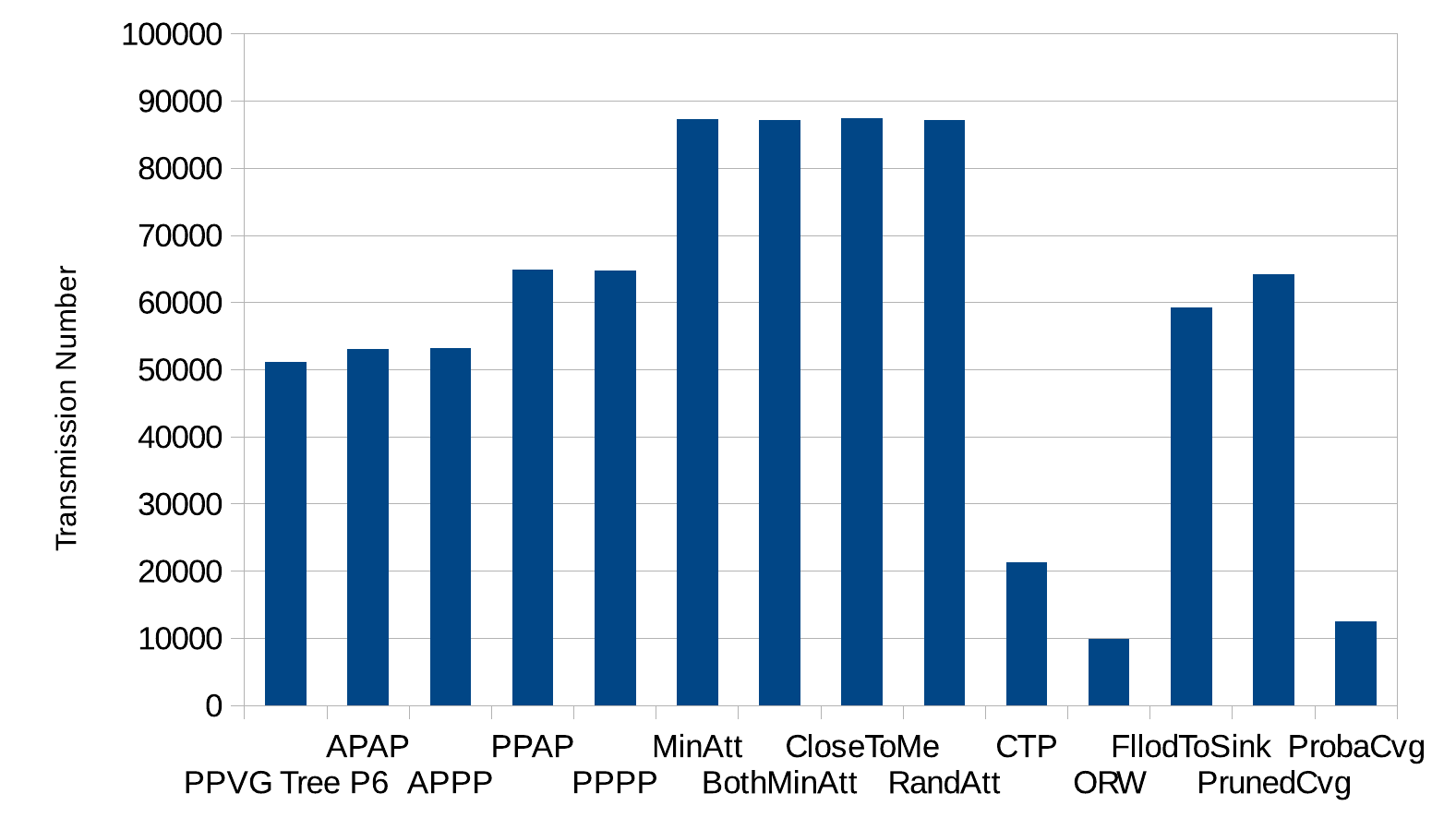, width=2.5in}
%\caption{Number of Transmissions in Posture 1- 4}
%\label{Fig40}
%\end{figure}

%\begin{figure}[!t]
\centering
\epsfig{file=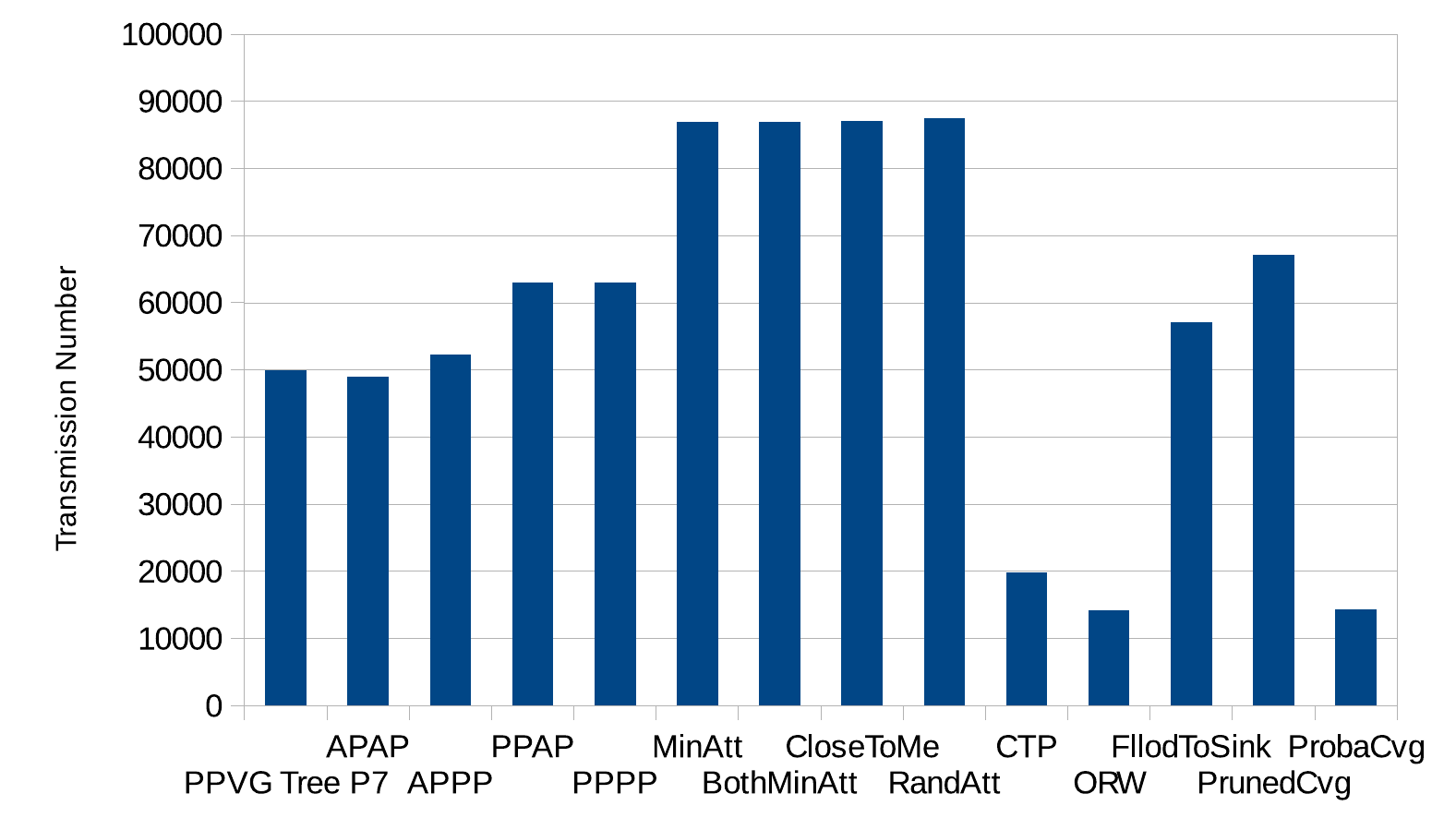, width=2.5in}
\caption{Number of Transmissions comparison among all the strategies in Posture 5-7}
\label{Fig41}
\end{figure}

From Figure \ref{Fig24} to Figure \ref{Fig41}, we give the comparisons of 1) Percentage of received messages (Figure \ref{Fig24} and Figure \ref{Fig27}) 2) Wrong Sequence rate (Figure \ref{Fig31} and Figure \ref{Fig34}) and 3) Number of Transmissions (Figure \ref{Fig38} to Figure \ref{Fig41}) at a rate of 10 packets par second (which is the maximum rate of actual WBAN medical applications) for each posture among all the strategies. All the strategies use the No-ACK Retransmission mechanism. CTP and ORW use the traditional ACK-based Retransmission mechanism. The presented results are the mean values over 10 simulation runs.

Our PPVG Tree strategy has the best Reception Rate for postures 3, 6 and 7; the second best for postures 1 and 2; the third best for postures 4 and 5. Since our PPVG Tree strategy is specify for different postures but no need additional rout-updating cost, our strategy give the a reception rate stable and relative high for every postures.

For Wrong-Sequence rate, our PPVG Tree strategy has the zero Wrong-Sequence rate for each posture, a good property of total order. For other strategies, there is always a possibility of appearing the Multi-Paths to the sink, so there are always the Wrong-Sequence.

For the number of transmissions, our PPVG Tree strategy always has the smallest transmission number compared with strategies who have comparable, like APAP, APPP and FloodToSink.

\section{Conclusion}
We first stressed existing WBAN convergecast strategies with rates of transmissions up to 500 packets per second in order to evaluate their capacity to be reliable and to ensure the total order message delivery. Our findings show that the performances of these strategies decrease dramatically (more than 90\% of the packets are dropped). Second, we proposed a new posture-centric model for WBAN and based on this  model we proposed a new mechanism for reliability. This mechanisms improves the existing strategies in terms of percentage of delivered messages. However, it does not help in increasing their total order reliability. We then proposed a new converge-cast strategy that outperforms WBAN dedicated strategies but also strategies adapted from DTN and WSN areas. Our extensive performance evaluations target: energy, percentage of message delivery and total order reliability (messages sent in a specific order should be delivered in that specific order). We used various rates of transmission and various human body postures. Our strategy ensures \emph{zero} packet order inversions. As future work we intend to investigate the fault-tolerance, security and privacy in WBAN.

\bibliographystyle{unsrt}%Used BibTeX style is unsrt
\bibliography{sample}

% You can push biographies down or up by placing
% a \vfill before or after them. The appropriate
% use of \vfill depends on what kind of text is
% on the last page and whether or not the columns
% are being equalized.

%\vfill

% Can be used to pull up biographies so that the bottom of the last one
% is flush with the other column.
%\enlargethispage{-5in}

% that's all folks
\end{document}